\documentclass[a4paper]{report}

\input{TEMP/Settings/packages}
\graphicspath{ {images/} }

\hypersetup{
    colorlinks,
    citecolor=blue,
    filecolor=blue,
    linkcolor=blue,
    urlcolor=blue
    }
    
    \makeatletter
\def\@makechapterhead#1{%
  {\parindent \z@ \reset@font
        \par\nobreak
        {\Huge \bfseries \thechapter\quad #1\par\nobreak}
        \par\nobreak
  }}

\makeatother

\definecolor{mygreen}{RGB}{28,172,0} 
\definecolor{mylilas}{RGB}{170,55,241}

\usepackage{chngcntr}
\usepackage[justification=centering]{caption}

\DeclareMathAlphabet{\mathcalligra}{T1}{calligra}{m}{n}
\DeclareFontShape{T1}{calligra}{m}{n}{<->s*[2.2]callig15}{}

\newcounter{questionctr}
\renewcommand{\thequestionctr}{\alph{questionctr}}

\newcounter{solutionctr}
\renewcommand{\thesolutionctr}{\alph{solutionctr}}

\newtcolorbox{solutionbg}{colback=white,colframe=black,arc=0pt,breakable}

\renewcommand*{\epsilon}{\varepsilon}

\newcommand{\thesistitle}{RMP screening by conducting structures in JOREK-CARIDDI}			
\newcommand{\yourname}{B.W. Slotema}				

\usepackage[T1]{fontenc}
\usepackage{times}
\usepackage{graphicx}
\usepackage{ifthen}
\usepackage{fancyhdr}

\parskip            \bigskipamount
\begin{document}

\begin{titlepage}
    \begin{figure}[h]
        \begin{subfigure}{0.5\textwidth}
            \includegraphics[width=0.8\linewidth, left]{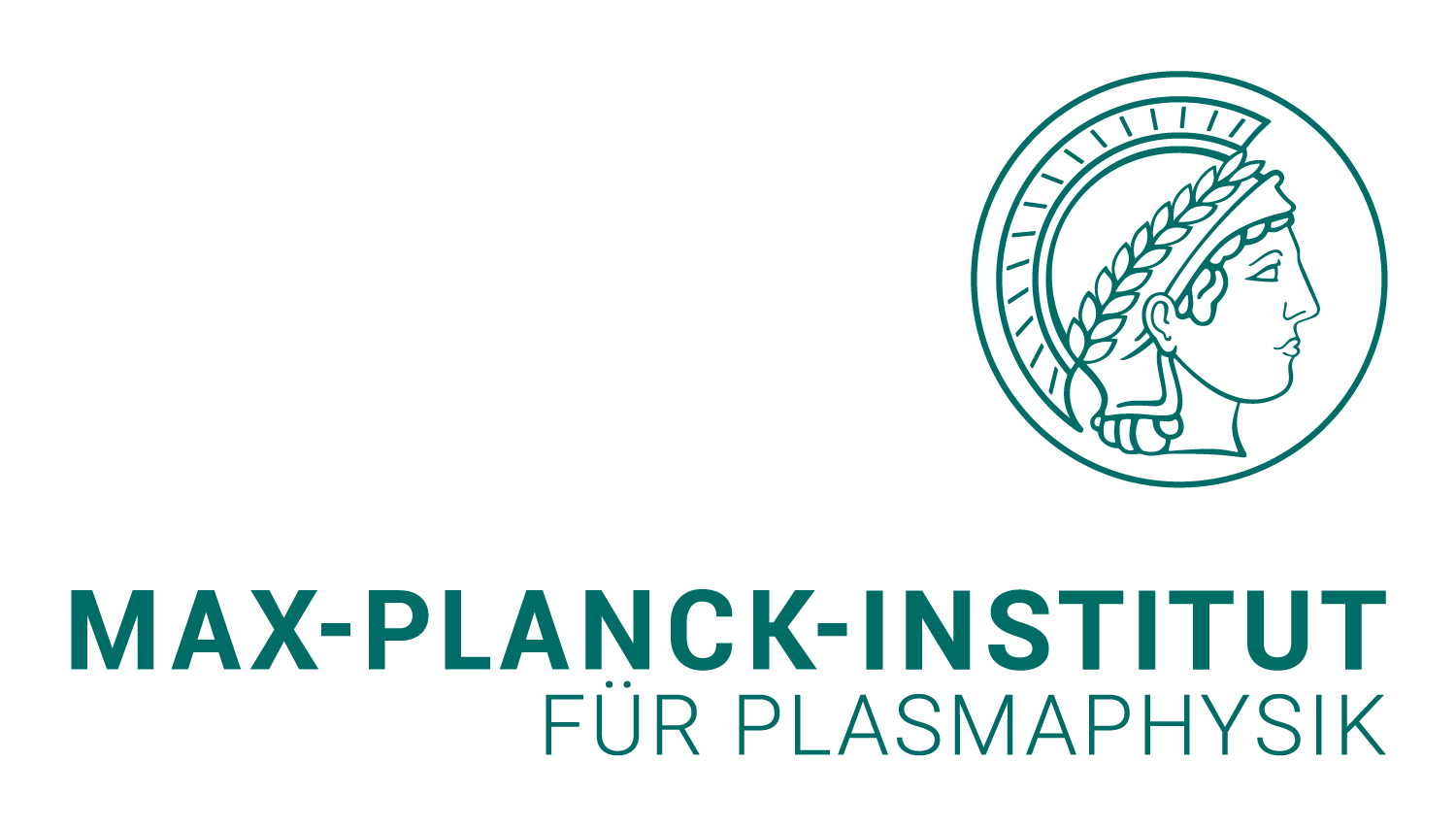} 
            \caption*{}
            \label{fig:subim1}
        \end{subfigure}
        \begin{subfigure}{0.5\textwidth}
            \includegraphics[width=0.8\linewidth, right]{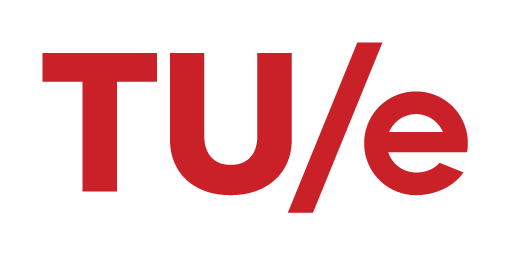}
            \caption*{}
            \label{fig:subim2}
        \end{subfigure}
    \end{figure}

    \vspace*{4cm}

    \begin{center}
        \LARGE{\textbf{\thesistitle}} \\
        \vspace*{0.5cm}
        
        \large{by} \\
        \vspace*{0.5cm}
        
        \Large{\textbf{\yourname}} \\
        \textit{1233729} \\
        \vspace*{2cm}
        
        \Large{Internship report}
    \end{center}
    
    \vspace{2cm}

    \centering

    \begin{tabular}{c c}
        \textbf{TU/e supervisor} & Prof. R.J.E. Jaspers \\
        \textbf{External supervisor} & Dr. M. Hoelzl, Group leader non-linear MHD\\
        \textbf{Daily supervisor} & V. Mitterauer \\
    \end{tabular}

    \vspace{1cm}

    \centering

    \begin{tabular}{c c}
        \textbf{Research conducted at:} & Max Planck Institute for Plasma Physics, Research unit MHD and fast particles \\
        \textbf{TU/e capacity group} & Science and Technology of Nuclear Fusion \\
        \textbf{Program} & Applied Physics \\
        \textbf{Study track} & Plasma \& Beams \\
        \textbf{Study load (ECTS)} & 15 \\
    \end{tabular}

    \thispagestyle{empty}
\end{titlepage}

\thispagestyle{empty}


\newpage
\chapter*{Abstract}

Edge localized modes (ELMs) are instabilities at the tokamak edge that can have short periodic outbursts of highly energetic particles and heat, which can severely damage the walls of a plasma reactor. Resonant magnetic perturbations (RMPs) are used to mitigate or eliminate ELMs from the plasma. A proper understanding of the interaction between the RMPs and other conducting structures in a tokamak is required to correctly operate a fusion reactor.
One important effect that can reduce the intensity of the RMP is screening, which is the result of eddy currents in conducting structures or a plasma that are induced by a time-varying magnetic field. The eddy current code CARIDDI was recently coupled with the magnetohydrodynamics (MHD) code JOREK, and is able to capture the behaviour of volumetric conducting structures that surround a plasma. The objective of this study is to characterize screening behaviour in the JOREK-CARIDDI coupling.\\

The analysis is divided in three parts. First, CARIDDI results are benchmarked against results from STARWALL, which is another JOREK extension that is able to capture interactions of (two-dimensional) conducting structures. The benchmarking is done with a static n=2 RMP, and it is found that CARIDDI and STARWALL show good agreement, with only small differences due to different representations of the RMP coils between the two codes. 
The second part of the analysis covers the screening of time-varying RMP fields by conducting structures in JOREK-CARIDDI, oscillating at frequencies spanning from 3 Hz to 10 kHz. The screening effect is less intense at low oscillation frequencies, which aligns with the expected behaviour, and saturates at frequencies above 100 Hz, at an amplitude of 28\%. The overall trends in screening behaviour align qualitatively with findings from a previous analysis. However, an additional contribution to the screening at frequencies above 1 kHz was not observed, because the simulation in this report does not include screening by the RMP coil casing.
Finally, a few tests were performed using a plasma configuration, in order to study the effect of plasma screening, including realistic plasma background flows. The results demonstrated the induction of eddy currents at rational surfaces within the plasma.\\
In summary, the results presented in this report demonstrated the ability of CARIDDI to qualitatively model RMPs that are induced by three-dimensional coils, as well as their screening by volumetric conducting structures and by the plasma. These results contribute to an improved understanding of the screening of RMP fields, which is relevant for several applications to the operation of fusion reactors. \\

\tableofcontents
\chapter{Introduction}
\label{chap:introduction}







The need for a renewable solution to address the growing global energy demand is widely acknowledged. Diverse sources, such as solar panels, wind turbines, and tidal turbines, have been developed to contribute to a more sustainable energy supply. However, a drawback is their reliance on specific conditions. For instance, solar panels are less effective on cloudy days, and wind turbines generate less energy during less windy days. \\

A form of sustainable energy that would not face this problem, is the use of fusion reactors. This type of reactor aims to emulate the processes that take place in the centre of stars. It makes use of the fact that the binding energy of a hydrogen isotope is higher than the binding energy of a helium atom. So, if one were to have an interaction where two $^2H$ isotopes merge (fuse) into a single helium atom, the total binding energy of the helium atom is less than that of the 2 original isotopes together. The excess energy in this reaction will be released in the form of kinetic energy of the resulting particles. The foreseen reaction candidate for fusion energy production is

\begin{equation}
    _{1}^{2}D + _{1}^{3}T \rightarrow _{2}^{4}He + _{0}^{1}n
    \label{eq:fusionwow}
\end{equation}

where $_{1}^{2}D$ is deuterium, $_{1}^{3}T$ is tritium, $_{2}^{4}He$ is helium and $_{0}^{1}n$ is a neutron. The reaction that is given in equation \ref{eq:fusionwow} releases a total amount of 17.6 MeV. Using one kilogram of deuterium-tritium would release a total energy of $3.4\times10^{14}$ J, which is 4 times as much energy as 1 kg of uranium would produce \cite{boom2013characterization}. \\

However, the reaction mentioned in equation \ref{eq:fusionwow} only happens when the deuterium and tritium atom are within the attraction range of the nuclear strong force, which is in the order of a femtometer ($10^{-15}$ meter). This is not easily achieved, since the atoms first need to overcome the Coulomb barrier, which is the electrical repulsion caused by the charges of their protons. Two protons need a total kinetic energy of $6\cdot 10^{-14} J$ to overcome this barrier in a head-on collision \cite{boom2013characterization}. In other words, the atoms need to have a very high temperature to have a sufficiently high kinetic energy. Using the fact that the average kinetic energy of a gas molecule at temperature $T$ is roughly equivalent to $3/2 k_B T$, where $k_B$ is Boltzmann's constant, one can deduce that a temperature of $3\cdot 10^{9}$ K is needed to overcome the Coulomb barrier. Luckily, due to tunneling and the fact that the velocity of a gas follows a distribution, this temperature reduces to approximately 0.12 $\cdot 10^9$ K. At these temperatures, the atoms are fully ionised and form a plasma. \\

To ensure effective reactions, the plasma needs to be contained. In fusion reactors, this responsibility is assumed by magnetic fields. Due to the Lorentz force, the motion of a charged particle perpendicular to a field line is very limited. To optimize confinement, the field lines should not intersect with the material walls. A widely used configuration in this context is the tokamak, which is shaped like a toroid and has magnetic coils. Figure \ref{fig:tokamak} provides a schematic setup of a tokamak. The magnetic field in a tokamak has a helical shape to minimize losses due to various magnetic effects.\\

An additional condition for the fusion condition, besides reaching a sufficiently high temperature to overcome the Coulomb barrier, is to have a sufficient number of interactions happening in the reactor. In other words, the collision frequency needs to be sufficiently high as well. Therefore, the particle number density also needs to be sufficiently high. Furthermore, the energy of the plasma should be confined long enough in order to reach ignition, which is the point where the energy that is produced by the fusion reactions is sufficient to heat the plasma. This is expressed by the minimum energy confinement time $\tau_E$. \\

To summarize, the plasma needs to have a sufficiently high temperature, have a high enough particle number density and the plasma energy should be confined well enough to sustain the fusion reaction. These three requirements must be met at the same time, and combine into a triple product called the Lawson criterion \cite{Lawson}:

\begin{equation}
    n T \tau_E > 5\cdot10^{21} m^{-3} keV s
    \label{eq:lawson}
\end{equation}

From the ideal gas law, it follows that an increase in number density and temperature leads to an increase in pressure. This pressure, called plasma pressure, needs to be balanced with the magnetic pressure to contain the plasma. The ratio of these two parameters is called $\beta$ and is given by:

\begin{equation}
    \beta = \frac{p_{plasma}}{p_{magnetic}} = \frac{n k_B T}{B^2/(2\mu_0)}
    \label{eq:plasmabeta}
\end{equation}

The $\beta$ parameter plays an important role in plasma control, since it describes the balance between the pressure of the plasma and the required magnetic field strength to contain the plasma. Ideally, the value for $\beta$ is slightly below 1.  \\

One very promising plasma confinement mode is called the H-mode, or high confinement mode, which will be treated in more detail in chapter \ref{chap:background}. One disadvantage of this confinement mode is that it features instabilities at the edge of the plasma, leading to so-called Edge Localized Modes (ELMs). These ELMs can be characterized in different types, depending on their size and frequency. The Type-I ELM has the potential to damage the plasma facing components in a tokamak. It was found that the Type-I ELM can be mitigated or suppressed by adding a small perturbation to the helical magnetic field that confines the plasma. This is done by so-called resonant magnetic perturbation (RMP) coils. \\

The magnetic perturbations that are generated by the RMP coils induce eddy currents in the surrounding structures, when the magnetic field of the RMP coil varies over time. These eddy currents, in turn, generate a magnetic field that opposes the initial magnetic field produced by the RMP coils, which is known as screening. This process has to be taken into account when configuring the currents carried by the RMP coils for ELM prevention, thereby impacting the overall efficiency of the tokamak.\\

This report aims to simulate the screening effect of passive conducting structures on the RMP magnetic field using the JOREK simulation code. In a recent development, JOREK was integrated with CARIDDI \cite{CARIDDI_coupling}, allowing for the simulation of eddy currents within volumetric conducting structures of the tokamak. The CARIDDI results will be benchmarked with STARWALL, another JOREK extension that is able to simulate wall effects using a simplified representation of conducting structures in a tokamak setup. STARWALL has been extensively used and tested, making it a valuable benchmark for comparison with CARIDDI, which is only recently coupled to JOREK and has had only limited uses to this point. The research question that will be answered in this report is:\\ 

\textit{Can the volumetric resistive wall code CARIDDI reproduce the screening of magnetic perturbation fields produced by RMP coils by passive conducting structures?}\\

The report is structured as follows: section \ref{chap:background} provides a brief overview of the theoretical foundations that are relevant to the investigation in this report. The configurations that were used for JOREK, STARWALL and CARIDDI are discussed in section \ref{chap:setup}. Next, Section \ref{chap:results} presents the results of the analysis. Section \ref{chap:discussion} provides suggestions for future research. In this section, the methods that were used in this report will also be evaluated. The report concludes in section \ref{chap:conclusion} by addressing the research question posed in the introduction and by providing a brief outlook.

\chapter{Background}
\label{chap:background}

\section{Tokamaks}
In the context of sustainable energy sources, fusion reactors have emerged as a promising candidate. Among the various reactor configurations, the tokamak stands out as the most commonly explored and investigated design for magnetic confinement fusion.
 A tokamak is a toroidally shaped device that has a strong helical magnetic field that is used for particle confinement. The helically shaped magnetic field is the result of combining a toroidal magnetic field with a poloidal magnetic field. The poloidal field component is typically an order of magnitude smaller than the toroidal component.\\

 \begin{figure}[!ht]
     \centering
     \includegraphics[width=0.7\textwidth]{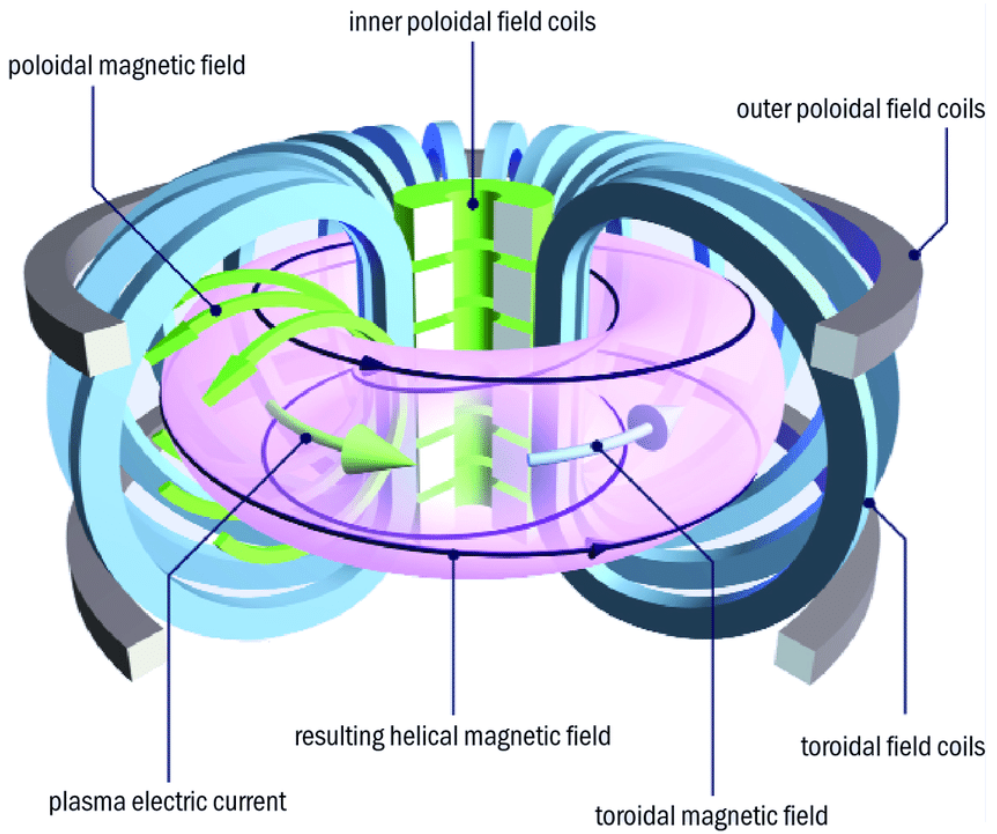}
     \caption{An illustration of the coils and the magnetic fields in a tokamak configuration. Figure taken from \cite{EUROFUSION}}
     \label{fig:tokamak}
 \end{figure}

 Figure \ref{fig:tokamak} displays a schematic drawing of a tokamak. The most relevant structures, as well as the ones that are most relevant for the research presented in this report, are:
 \begin{itemize}
     \item \textbf{Toroidal field coils}: These coils are responsible for producing the magnetic field in the toroidal direction. The coils are evenly distributed in the toroidal direction. Present-day tokamaks commonly use copper coils. However, for future devices like ITER, the use of superconducting materials will be required to efficiently sustain a strong magnetic field strength over a long time.
     \item \textbf{Poloidal field coils}: The main component of the poloidal field is produced by driving a current through the plasma in the toroidal direction, called the plasma current. The inner poloidal field coils induce the plasma current, while the outer poloidal field coils produce a vertical magnetic field, which is used for plasma shaping and position control.
     \item \textbf{Passive stabilizing loops (PSL)}: The PSL consists of two loops, located on the upper and lower side of the plasma, that are connected with each other. The PSL serves to stabilize the plasma against self-amplifying vertical movement. In ASDEX Upgrade, the PSL can partially shield the RMP fields, since the RMP coils are located closely to the PSL.
     \item \textbf{Resonant magnetic perturbation coils (RMP coils)}: These coils are located inside the vacuum vessel and provide a small perturbation to the helical field of the plasma. This is done to suppress Edge Localized Modes (ELMs), as treated in section \ref{sec:ELM}. 
     \item \textbf{Tokamak wall}: The wall of a tokamak serves as a boundary that separates the plasma from the surrounding environment. It is constructed from materials that are able to withstand extreme conditions and also serves to protect the plasma from outside influences. Furthermore, it can interact with the plasma due to its finite conductivity, and is able to influence the stability of the plasma or influence the impact of the external perturbations provided by RMP coils.
     \item \textbf{Divertor}: The divertor is a region where impurities and $_2^4He$ atoms are guided to and absorbed. Divertors are commonly placed at the bottom of the tokamak. The materials of the divertor region are designed to withstand high particle and heat fluxes.
 \end{itemize}


\section{Magnetic confinement} 

As stated in the introduction, achieving fusion reactions requires having a high plasma temperature, as dictated by the Lawson criterion. In a tokamak, this is achieved by externally heating the plasma. Two commonly used methods for this purpose are neutral beam injection \cite{manheimer1985tokamak} and microwave heating \cite{takahashi1977icrf}. In a reactor, external heating would be applied until ignition is achieved, which is the point where the fusion reactions provide a sufficient amount of energy to maintain the plasma temperature (without the external heating). To minimize the required amount of external heating and reach ignition more easily, it is essential to keep the energy losses at a minimum.\\

 The particle transport along a magnetic field line is multiple orders of magnitude faster than perpendicular to a magnetic field line. Therefore, the magnetic field lines within a tokamak wrap around the torus, to eliminate direct parallel loss paths along the wall of the tokamak \cite{freidberg_2007}. The remaining energy losses in a tokamak are mainly caused by heat and particle diffusion, e.g. induced by turbulence caused by small scale instabilities in the plasma. \\ 

\begin{figure}[ht!]
   \centering
    \includegraphics[width=0.6\linewidth]{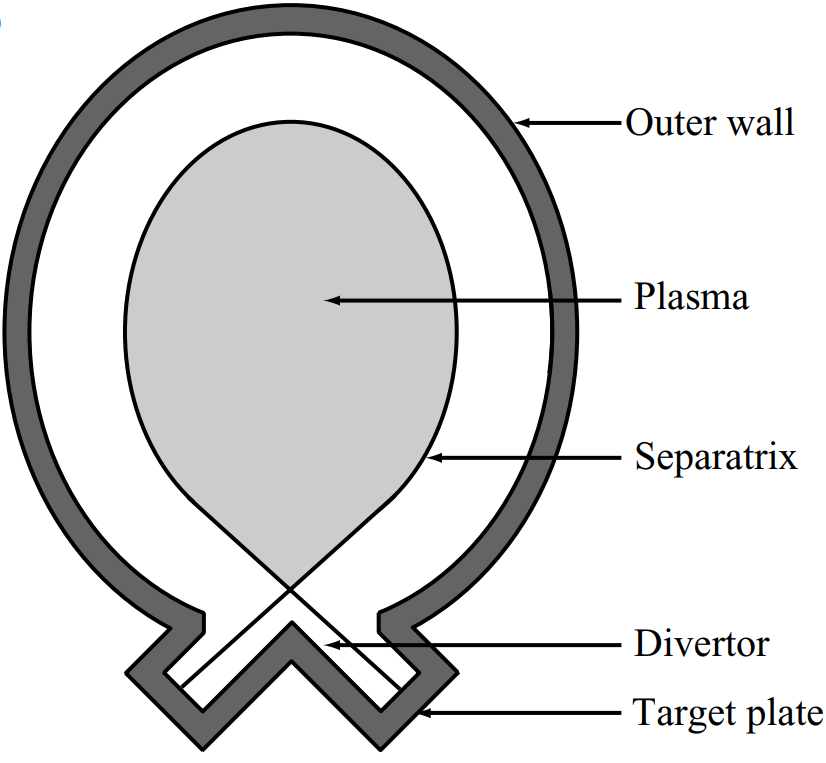}
\caption{Schematic representation of the x-point configuration in a tokamak. Taken from \cite{freidberg_2007}. Reproduced with permission of The Licensor through PLSclear. © J. Freidberg 2007}
\label{fig:divertor}
\end{figure}

A commonly used tokamak setup is the x-point configuration, which is shown in figure \ref{fig:divertor}. This setup allows for more flexible shaping of the plasma, which can significantly improve confinement time. The configuration consists of a region with open flux surfaces (scrape off layer or SOL) and a region of closed flux surfaces. The flux surface that separates both regions is called the separatrix. Furthermore, there is a point at which the poloidal field vanishes, which is referred to as the x-point. The particles in the SOL are guided towards the divertors at the bottom of the tokamak.

\begin{figure}[h!]
    \centering
    \includegraphics[width=0.8\textwidth]{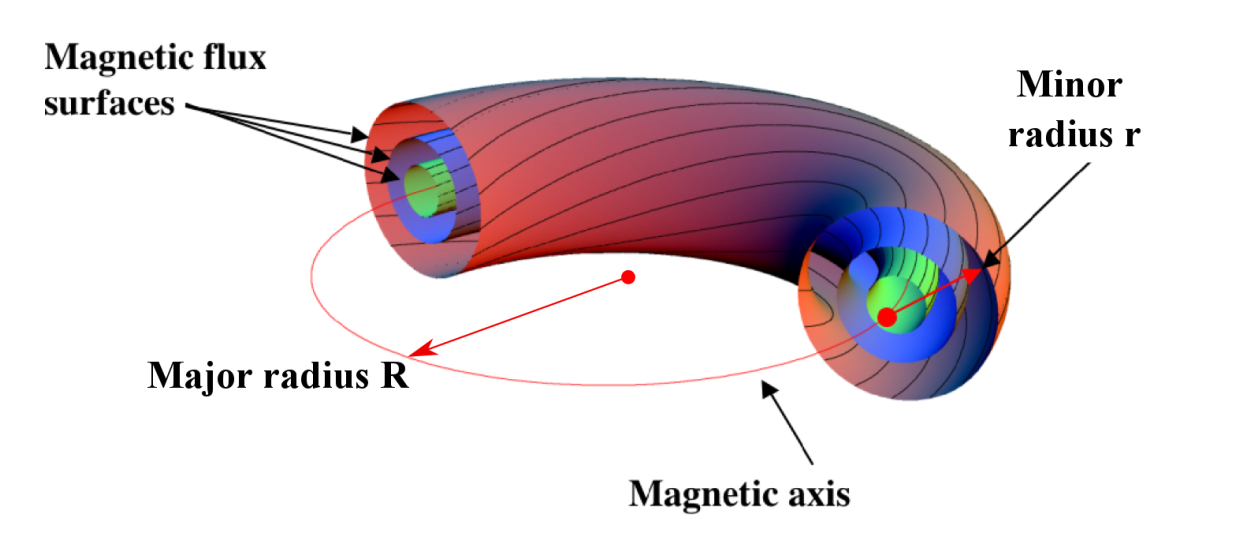}
    \caption{The magnetic flux surfaces inside a tokamak. Each colour represents a different flux surface. Figure taken from \cite{JAVIERthesis}}
    \label{fig:magsurf}
\end{figure}

 Within the closed flux surface region of a plasma, the magnetic field lines combine into nested structures known as magnetic flux surfaces, as illustrated in figure \ref{fig:magsurf}. The nested flux surfaces within a toroidal configuration fulfil the force-balance equation of magnetohydrodynamics (MHD) (as discussed in section \ref{sec:MHD}). The force-balance equation solves equilibria in MHD, indicating that the pressure, temperature and density of the plasma are constant along a flux surface. Particles in a plasma are confined to a flux surface, and can only move to another by means of collision or by turbulent or neoclassical transport. The magnetic flux, used as a radial coordinate, is given by $\psi=A_\phi/R$ and is constant on a flux surface. \\
 
 Within a tokamak, the magnetic field lines are nearly static during normal operation. Therefore, they serve as a reliable way to describe the magnetic geometry of the tokamak.  An essential parameter in this context is the safety factor q, which is given by $q=m/n$ (or equivalently $q=d\phi /d\theta$), where m is the number of turns in the toroidal direction and n the amount of turns in the poloidal direction. Furthermore, $\phi$ is defined as the angle in the toroidal direction and $\theta$ is defined as the angle in the poloidal direction. Essentially, q describes the amount of toroidal turns that are needed to complete one single poloidal turn. High values of the safety factor at the edge provide more stable plasmas. When the safety factor can be expressed by a low-order rational number, the magnetic field lines close in on themselves, indicating that it returns to the same point after a finite number of turns. Rational field lines are more prone to plasma instabilities, because they are brought to resonance more easily. \\


\subsection{H-mode}
\begin{figure}[!ht]
    \centering
    \includegraphics[width=0.4\textwidth]{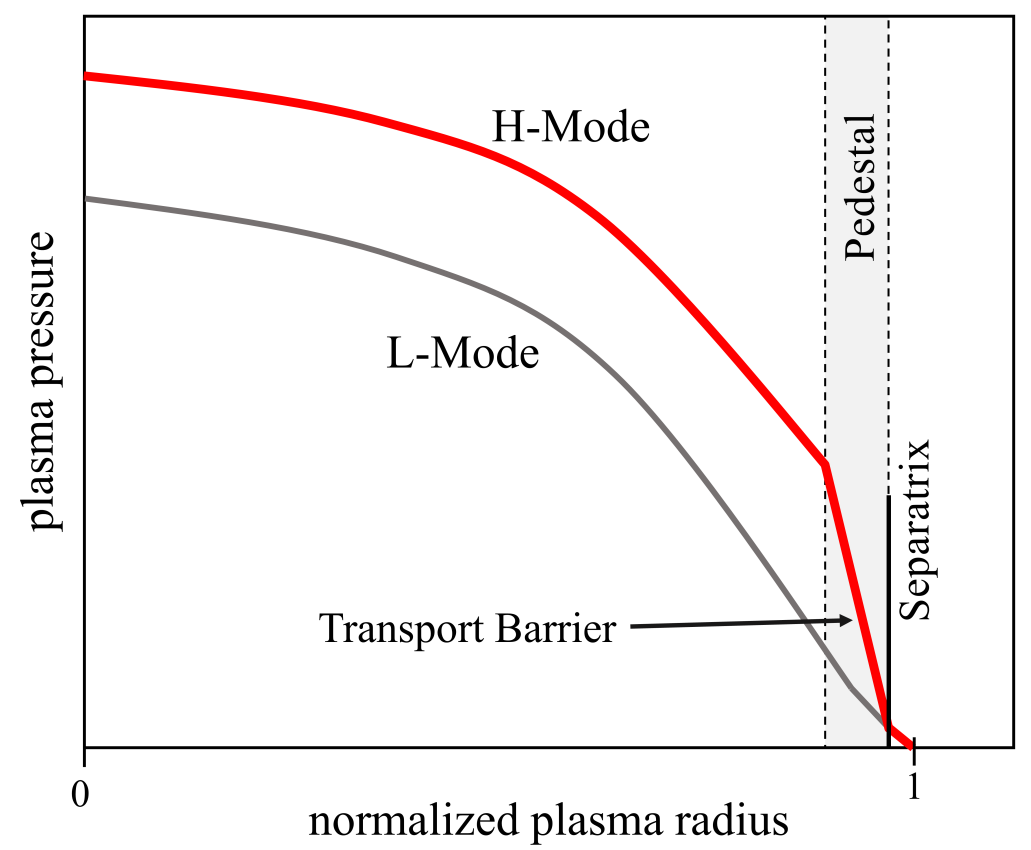}
    \caption{Typical pressure profiles observed in L- and H-mode phases. The L-mode indicates the low-confinement mode, which is not treated in this report. Figure taken from \cite{mitterauer2021non}. }
    \label{fig:pedestalpressure}
\end{figure}

When the heating power in a tokamak is increased above a certain threshold, the plasma undergoes a transition into a state of improved confinement, commonly referred to as the high-confinement mode or H-mode \cite{Wagner82}. This state leads to a higher temperature and density at the plasma core. The improved confinement arises from a thin layer at the plasma edge, known as the edge transport barrier (ETB). Within the ETB, particle and energy transport experience a notable reduction, and the pressure exhibits a steep gradient, creating a region known as the pedestal, which is shown in figure \ref{fig:pedestalpressure}. One requirement of the H-mode is that the plasma temperature at the edge remains high, which is most easily obtained in divertor tokamaks. As a result, divertor tokamaks are preferred for future plasma reactors \cite{Stangeby_1990}. 

\section{Edge localized modes}
\label{sec:ELM}

One disadvantage of the H-mode is that it is susceptible to instabilities in the plasma, leading to edge localized modes (ELMs). These ELMs cause a transient collapse of the ETB, which leads to short periodic outbursts that expel energy and particles from the confined plasma, which can damage the wall of the tokamak. The collapse of an ELM typically lasts only for a timescale in the order of a few hundred $\mu s$.\\
After the ELM crash, there is a recovery phase where the density and temperature gradients build up again. The duration of this recovery phase determines the frequency of the ELMs.\\

\begin{figure}[!ht]
    \centering
    \includegraphics[width=0.5\linewidth]{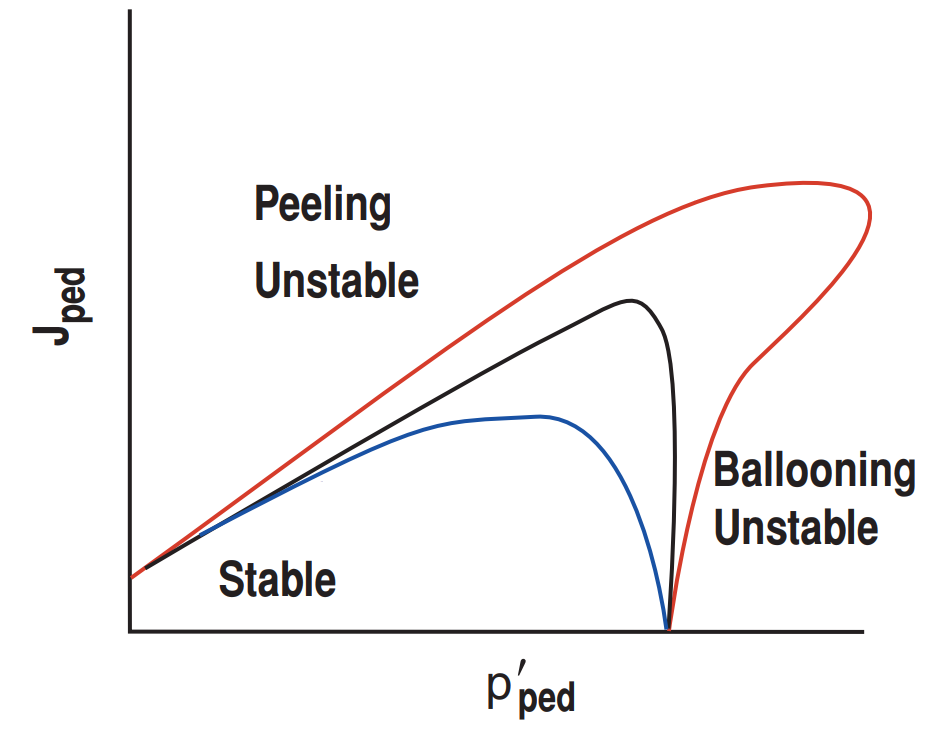}
    \caption{Peeling-ballooning mode stability regions as a function of pedestal pressure gradient ($\mathbf{p_{ped}'}$ and pedestal current $J_{ped}$. Figure taken from \cite{MBongard}}
    \label{fig:PBstability}
\end{figure}

According to ideal MHD, the unstable modes that trigger ELMs are driven by both the pressure gradient and the current at the plasma edge. The so-called ballooning modes are caused by the pressure gradient. For more information about ballooning modes, the reader is referred to \cite{Taylor_2012}. On the other hand, peeling modes are instabilities that are caused by the current at the plasma edge. The reader is referred to \cite{PeelingMode} for more information about the peeling mode. The modes couple, creating the so-called peeling-ballooning mode. Figure \ref{fig:PBstability} illustrates the plasma stability dependence on the pressure gradient and edge current. \\

Type-I ELMs have the highest amount of energy release and can expel up to 10\% of the total plasma energy \cite{JAVIERthesis}, and can damage the surrounding structures of the tokamak, and should therefore be kept under control. For the remainder of this report, the term "ELMs" will refer exclusively to Type-I ELMs, unless explicitly stated otherwise. \\

In present-day tokamaks, the ELM heat loads are still acceptable. However, for bigger tokamaks such as ITER, the flux will be too high and the ELMs need to be suppressed or mitigated \cite{ITERdamage}. Several methods have been devised to mitigate or suppress ELMs. These methods include pellet injection \cite{Lang_2012}, vertical kicks \cite{VerticalKicks} and resonant magnetic perturbations (RMPs) \cite{DIII_RMP}.

\section{Resonant Magnetic Perturbations}
Resonant Magnetic Perturbations (RMPs) are small non-axisymmetric perturbations to the magnetic field, which have been successfully used for ELM control in many tokamaks, including  DIII-D \cite{DIII_RMP}, EAST \cite{EAST_RMP}, MAST \cite{MAST_RMP}, KSTAR \cite{KSTAR_RMP} and ASDEX Upgrade (AUG) \cite{AUG_RMP}. However, the mechanism that underlies ELM suppression by RMPs is not yet fully understood.\\

\begin{figure}[!ht]
    \centering
    \includegraphics[width=0.5\textwidth]{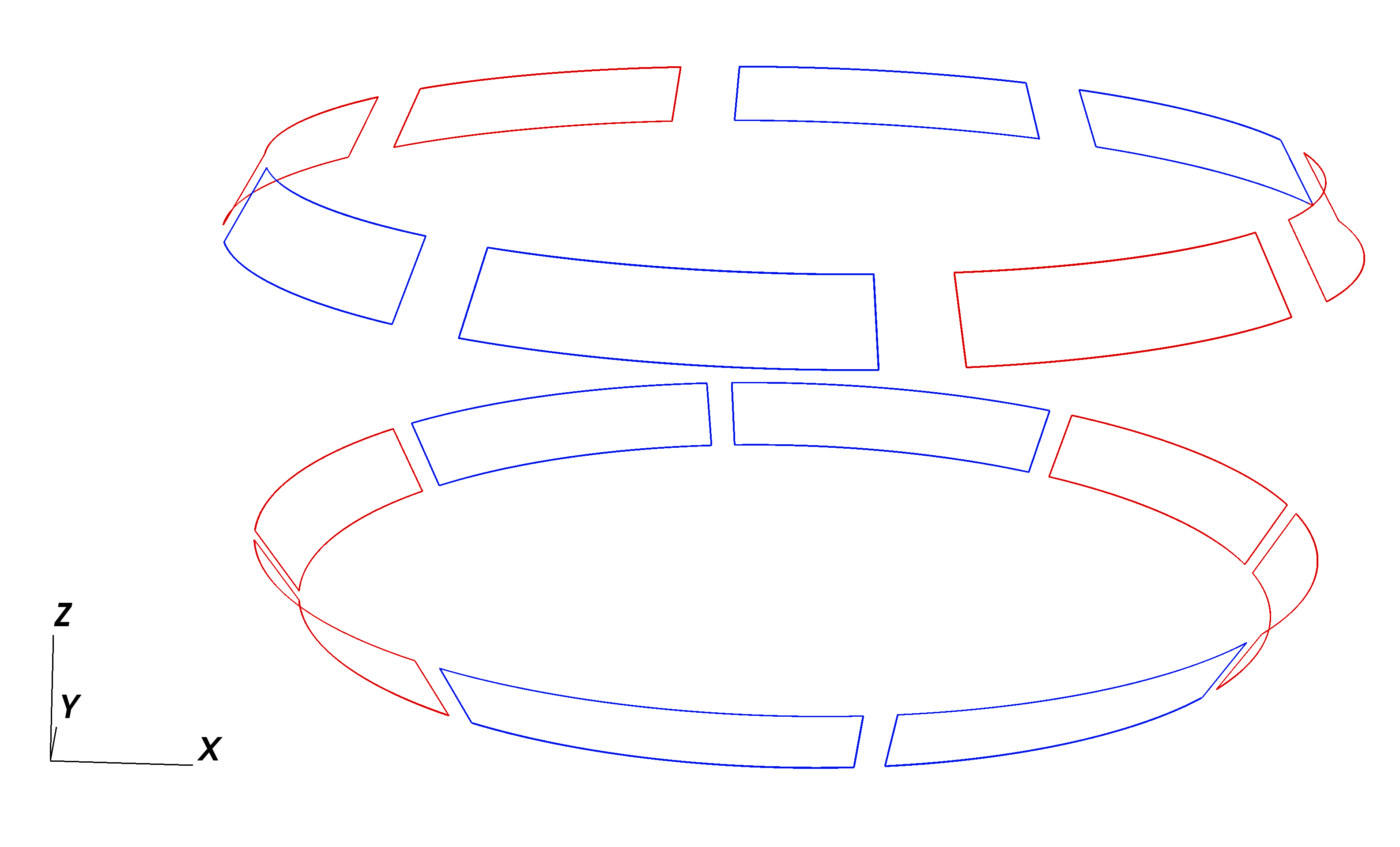}
    \caption{Schematic of the RMP coil configuration in AUG for an n=2 perturbation. The red and blue colours correspond to positive and negative coil current, respectively.}
    \label{fig:coilsss}
\end{figure}

 RMPs are generated using external coils known as RMP coils, which are strategically positioned inside the vacuum vessel of the tokamak. These coils are typically organized in a ring configuration at a specific height, often located near the top and bottom of the tokamak. Figure \ref{fig:coilsss} shows an example of a RMP configuration.  RMP coils allow for the application of diverse magnetic fields by controlling the currents flowing through each individual coil. For example, the currents in figure \ref{fig:coilsss} produce a magnetic field that oscillates twice along the circumference of the tokamak. This is referred to as an n=2 RMP, where n represents the toroidal mode number. With 16 coils, one can achieve a perturbation with a toroidal mode number of 1, 2, 4 or 8. Other toroidal mode numbers are also possible, depending on the configuration and amount of RMP coils in a tokamak. The magnitude of an RMP field is approximately $\delta \mathbf{B}/\mathbf{B} \sim 10^{-3}-10^{-4}$. \\ 

\subsection{RMP screening}

The magnetic field that is generated by RMP coils is usually not time-varying, except during the ramp-up phase, which is directly after the coils are switched on. Furthermore, it is possible to produce a time-varying RMP by driving a time-dependent coil current through the RMP coils. In accordance with Faraday's law, a time-varying magnetic field induces an electric field:

\begin{equation}
    \vec{\nabla}\times\vec{\mathbf{E}} = -\frac{\partial\textbf{B}}{\partial t}
    \label{eq:Fara}
\end{equation}

\begin{figure}
    \centering
    \includegraphics[width=0.7\textwidth]{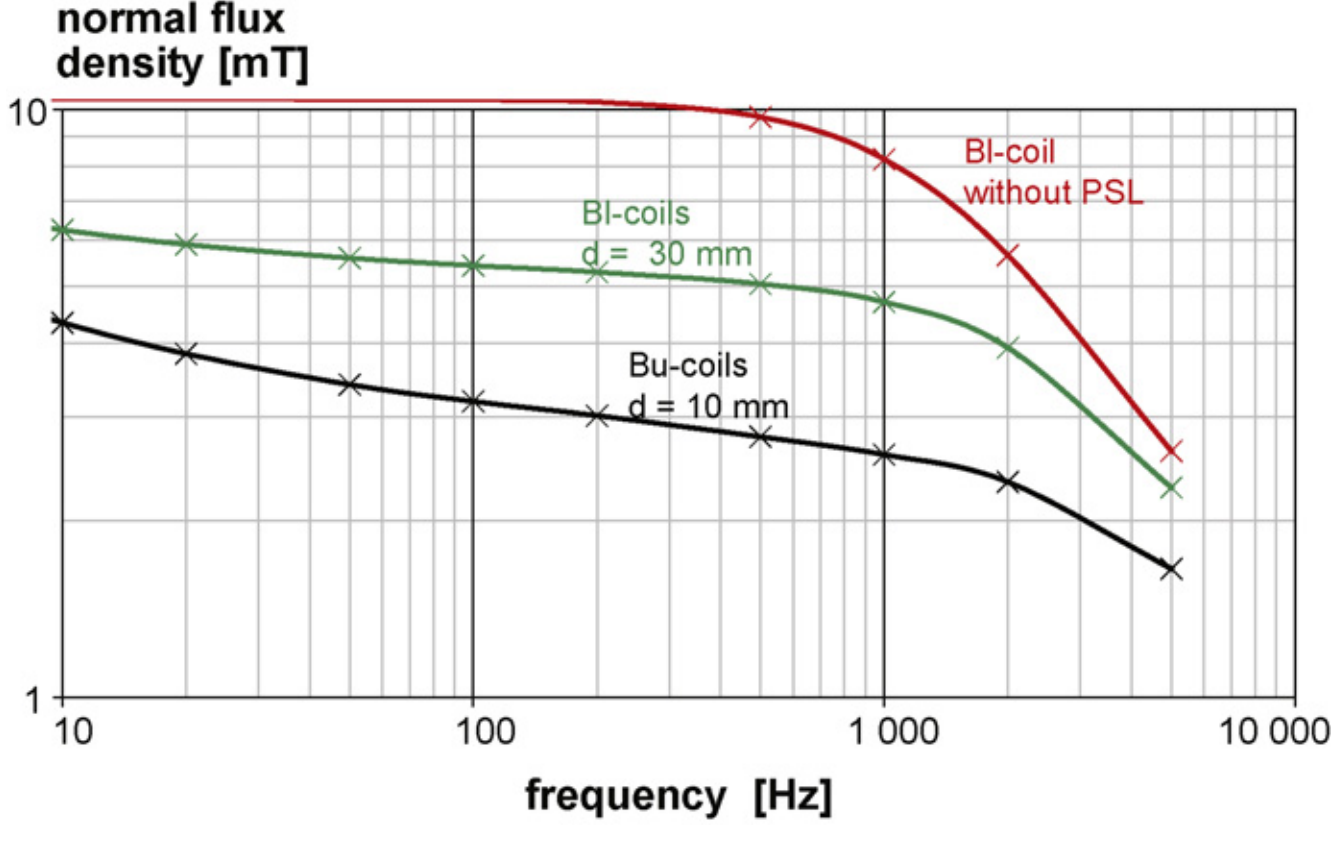}
    \caption{Magnetic flux density produced by RMP coils with a coil current of 1 kA as
a function of coil current oscillation frequency for varying distance between coil and PSL. The Bl and Bu coils correspond to the lower and upper RMP coil ring, and d refers to the mounting distance to the PSL. Taken from \cite{updatedcoils}}
    \label{fig:freqscreen}
\end{figure}

This induced electric field subsequently produces eddy currents within the conducting structures surrounding the RMP coils. Lenz's law dictates that the direction of these eddy currents opposes the change in magnetic flux that was responsible for its initiation. Consequently, the induced current gives rise to a magnetic field, given by Ampere's law. This resultant magnetic field undergoes destructive interference with the magnetic field emanating from the RMP coils, a phenomenon known as screening, which will be the main subject of the research that is presented in this report. The screening effect is more intense if the magnetic field changes more rapidly, as shown in figure \ref{fig:freqscreen}.\\

An additional contributor to the screening of the RMP magnetic field is the plasma itself, which can significantly influence the RMP field in the plasma and also affect the penetration of the RMP field in the plasma. This screening is mainly caused by currents that are induced in the plasma. In the pedestal region, RMP screening is mainly caused by $\Vec{E}\times\vec{B}$ rotation and diamagnetic electron rotation, also referred to as rotational screening. For rotational screening, the electrons move across the field lines at the resonant surfaces. For an electron with rotational motion, an RMP that is static in the lab frame is seen as time-varying in the electron fluid frame. According to equation \ref{eq:Fara}, this will induce an electron current that will lead to screening effects. For a more detailed explanation, the reader is referred to \cite{plasmascreeningI} \cite{plasmascreening_II}. 

\section{ASDEX Upgrade}

ASDEX Upgrade (AUG), located in Garching, Germany, serves as a tokamak for fusion research. It is the largest tokamak in Germany, and one of the leading fusion experiments worldwide \cite{AUG}. The experimental configuration of ASDEX-Upgrade has been adopted as the basis for the simulations presented in this report. This report will use results from AUG as input for plasma simulations. The motivation for selecting this particular experiment lies in the accessibility of data, attributed to the involvement of the Max Planck Institute for Plasma Physics in this project. Furthermore, the magnetic geometry is similar to that of ITER, where AUG is close to a scaled-down version of ITER \cite{AUG}. A comprehensive overview of the configuration of the ASDEX-Upgrade tokamak can be found in \cite{AUG_setup}.\\


The configuration of the coils that are used to create the resonant magnetic perturbation (RMP) is displayed in figure \ref{fig:coilsss}. In total, AUG has 16 RMP coils that are placed within the plasma vessel, where 8 coils are placed in the negative z plane and 8  coils in the positive z plane. 

In the AUG experiment, the coils are powered with a coil current up to 6.5 kAt (kiloampere-turn), and are configured to produce a magnetic field with an n=2 component in the toroidal direction \cite{AUG_rmp_schematic}, though configurations with n = 1 and n = 4 are also possible \cite{AUGn14}.

\section{Magnetohydrodynamics}
\label{sec:MHD}
Magnetohydrodynamics (MHD) is a theoretical framework that is used to describe (fluid) dynamics of electrically conducting fluids. Its equations are primarily derived from electromagnetism and fluid mechanics and describe the evolution of the magnetic potential (\textbf{A}), the mean plasma velocity (\textbf{V}), the total plasma density ($\rho$) and total pressure ($p$) of the plasma. \\
MHD has proven to be effective in anticipating the onset of certain plasma instabilities. In the context of fusion reactors, MHD serves as a valuable tool for modelling plasma dynamics. These models help understand the fundamental processes that take place inside a reactor plasma.\\

The derivation of MHD theory is omitted in this paper, but the reader is referred to \cite{freidberg_2007} \cite{MHD} for a more complete description of MHD. Various versions of MHD exist, depending on the assumptions made about the system. For ideal MHD, the following assumptions are made:
\begin{itemize}
    \item The plasma is strongly collisional, such that the collision timescales are much smaller than all the other timescales in the system. Consequently, the particle distributions are close to Maxwellian
    \item The resistivity that is due to the plasma collisions is small, such that the timescale for magnetic diffusion must be much larger than any other timescale in the system
    \item The length scale of any parameter of interest is much larger than the ion skin depth and the Larmor radius perpendicular to the magnetic field, such that the individual particle effects can be averaged out, and a fluid description is appropriate. Furthermore, the wave-particle interactions that lead to Landau damping are not significant over the length scales of interest
    \item The time scales of interest are much longer than the ion gyration time, indicating that the system is smooth and evolves slowly
\end{itemize}

The fundamental equations of ideal MHD are given in equation \ref{eq:MHD}.

\begin{subequations}
\begin{align}
    & \text{mass conservation:} && \frac{\mathrm{d} \rho}{\mathrm{d} t}+\rho \nabla \cdot \mathbf{v} = 0 \label{eq:mass} \\
    & \text{momentum conservation:} && \rho \frac{\mathrm{d} \mathbf{v}}{\mathrm{d} t} = \mathbf{J} \times \mathbf{B}-\nabla p \label{eq:momentum} \\
    & \text{Ohm's law:} && \mathbf{E}+\mathbf{v} \times \mathbf{B} = 0 \label{eq:ohm} \\
    & \text{energy conservation:} && \frac{\mathrm{d}}{\mathrm{d} t}\left(\frac{p}{\rho^\gamma}\right) = 0 \label{eq:energy} \\
    & \text{Maxwell:} && \nabla \times \mathbf{E} = -\frac{\partial \mathbf{B}}{\partial t} \label{eq:maxwell1} \\
    &&& \nabla \times \mathbf{B} = \mu_0 \mathbf{J} \label{eq:maxwell2} \\
    &&& \nabla \cdot \mathbf{B} = 0 \label{eq:maxwell3}
\end{align}
\label{eq:MHD}
\end{subequations}

Here, the derivative $\frac{d}{dt} = \frac{\partial}{\partial t} + \vec{v}\cdot\vec{\nabla}$. Furthermore, $\mathbf{J}$ indicates the current density, defined as $\mathbf{J}=e(n_i\mathbf{u}_i-n_e\mathbf{u}_e$), $\mathbf{B}$ is the magnetic field of the plasma, $\gamma$ is the adiabatic constant (equal to 5/3) and $\mu_0$ is the vacuum permittivity constant. 

The MHD model that was used in this report was visco-resistive reduced MHD (RMHD) \cite{RMHD}, which is applicable for the problems at hand. The most important feature of RMHD is that it removes MHD phenomena that occur on very fast timescales and that it reduces the number of variables as compared to full MHD. These properties make RMHD an appealing candidate for simulations, as it reduces the amount of computational resources needed for a simulation. RMHD contains some additional assumptions, most importantly the assumption that the toroidal magnetic field is static\cite{ReviewPaper_2021}.\\


The time-scales for MHD are in the order of the Alfvén time, which is given by equation \ref{eq:Alfven}

\begin{equation}
    \tau_A = \frac{a}{v_A}
    \label{eq:Alfven}
\end{equation}

where $a$ is the minor radius of the tokamak and $v_A$ is the Alfvén velocity, given by $v_A = |\mathbf{B}|/\sqrt{\mu_0 \rho}$, with \textbf{B} the magnetic field in the tokamak, $\mu_0$ the magnetic permeability and $\rho$ the plasma density. For most tokamaks, the Alfvén time is in the order of microseconds.

\section{JOREK}

Due to the complexity of the equations that underlie MHD, a comprehensive understanding of all the model's properties is achievable only through numerical simulations. An example of such simulation software is JOREK, a non-linear code that is designed to simulate and study plasma instabilities in divertor tokamaks \cite{ReviewPaper_2021}. \\

\begin{figure}[h!]
    \centering
    \includegraphics[width=0.7\textwidth]{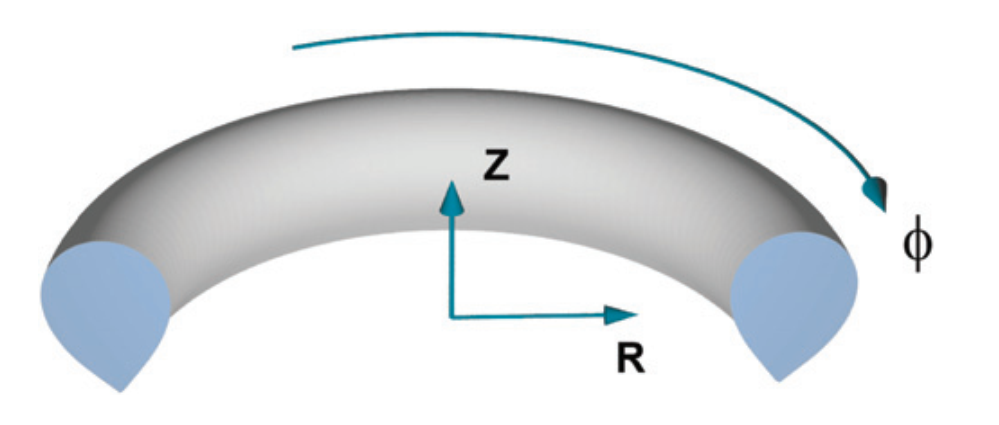}
    \caption{The coordinate system used in JOREK. Figure taken from \cite{ReviewPaper_2021}}
    \label{fig:coords}
\end{figure}

JOREK uses a cylindrical coordinate system, which is shown in figure \ref{fig:coords}. The physical variables inside the reactor volume are decomposed in terms of a 2D finite element expansion combined with a toroidal Fourier expansion. The order of this expansion can be configured using the number of toroidal harmonics to be used in the simulation. Furthermore, the periodicity of these modes can be configured, which guarantees periodicity after $\frac{1}{n_{period}}$ of a full toroidal turn. JOREK provides a selection of different models, so that a variety of different cases can be simulated efficiently and accurately. \\

Using the differential formulations that are used in JOREK, one of the required boundary conditions is to know the tangential magnetic field at the boundary of the computational domain ($\mathbf{B}_{tan} = \mathbf{B}\times\hat{\textbf{n}}$) \cite{CARIDDI_coupling}.\\

\section{Modelling of conducting structures}
\label{sec:SWCD}
To obtain an accurate description of the plasma dynamics, it is important to have a good understanding of the interactions of the plasma with the conducting structures surrounding the plasma. JOREK was integrated with the STARWALL code, to simulate the plasma dynamics in the presence of conducting structures, such as resistive walls and coils surrounding the tokamak.  STARWALL assumes the thin wall approximation in order to simplify the numerical calculation \cite{STARWALL_coupling}. In STARWALL, the response of 3-dimensional conducting structures to the magnetic perturbation is obtained by solving the vacuum magnetic field equation outside the computational domain of JOREK \cite{STARWALL_coupling}. A recent extension allows the use of non-axisymmetric active coils in STARWALL \cite{STARWALL_freeboundary}, which is crucial for modelling the screening effects of conducting structures on RMP coils.\\

Much like the JOREK-STARWALL coupling, the development of JOREK-CARIDDI was motivated by the need to understand the impact of conducting structures surrounding the plasma \cite{CARIDDI_coupling}. While STARWALL is limited to using thin walls, CARIDDI uses a fully volumetric 3D model for conducting structures \cite{CARIDDI_OG} and models the eddy currents within those structures. This coupling makes it possible to realistically model the 3D interactions between a plasma and the surrounding wall structures, making it possible to accurately simulate the screening of the RMP field, e.g. by the PSL. JOREK-CARIDDI was only recently coupled\cite{CARIDDI_coupling} \cite{CARIDDI_Nina}. This report will present one of the first user cases of the JOREK-CARIDDI coupling.

\chapter{JOREK setup} 
\label{chap:setup}

\begin{figure}[h!]
    \centering
    \includegraphics[width=0.9\textwidth]{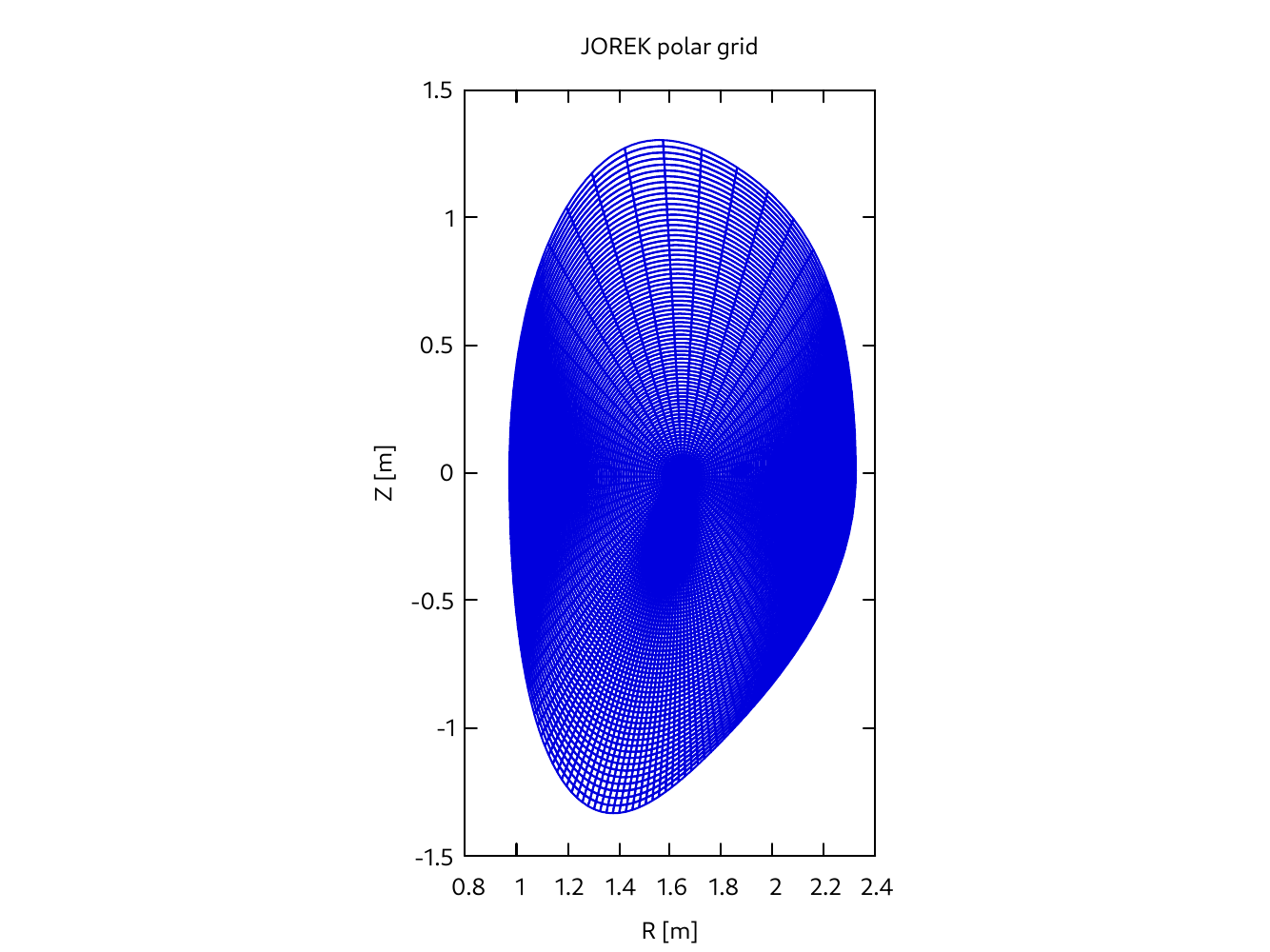}
    \caption{Illustration of the grid that was used in the simulations presented in this report}
    \label{fig:polargrid}
\end{figure}

The same basic JOREK setup is used for all analyses presented in this report. Appendix \ref{sec:params} provides an overview of all relevant parameters that were modified in JOREK for each analysis. The toroidal modes n=0,1,…,6 were used (JOREK counts the sine and cosine separately). 
The base grid for all simulations was a polar grid, with 80 radial grid points and 90 poloidal grid points. Figure \ref{fig:polargrid} displays the grid that was used in this report.\\ 

The positioning of the coils was based on ASDEX Upgrade (AUG), with one ring of 8 RMP coils in the negative z-plane and one ring of 8 RMP coils in the positive z plane. The coil currents were driven in such a way that an n=2 RMP was produced for all results written in this paper, unless stated otherwise. The values of the coil currents are shown in table \ref{tab:coilcurrs} in the Appendix.

\section{Vacuum state}
The focus of this report will be the screening of the magnetic fields induced by RMP coils by the surrounding conducting structures. Thus, a vacuum is approximated within the plasma vessel to eliminate plasma screening effects. The vacuum state will be 'created' by cooling down the initial plasma and by diffusing the particles inside the plasma vessel. 

\begin{figure}[!h]
    \centering
    \begin{subfigure}[b]{0.49\textwidth}
        \centering
        \includegraphics[width=1.1\textwidth]{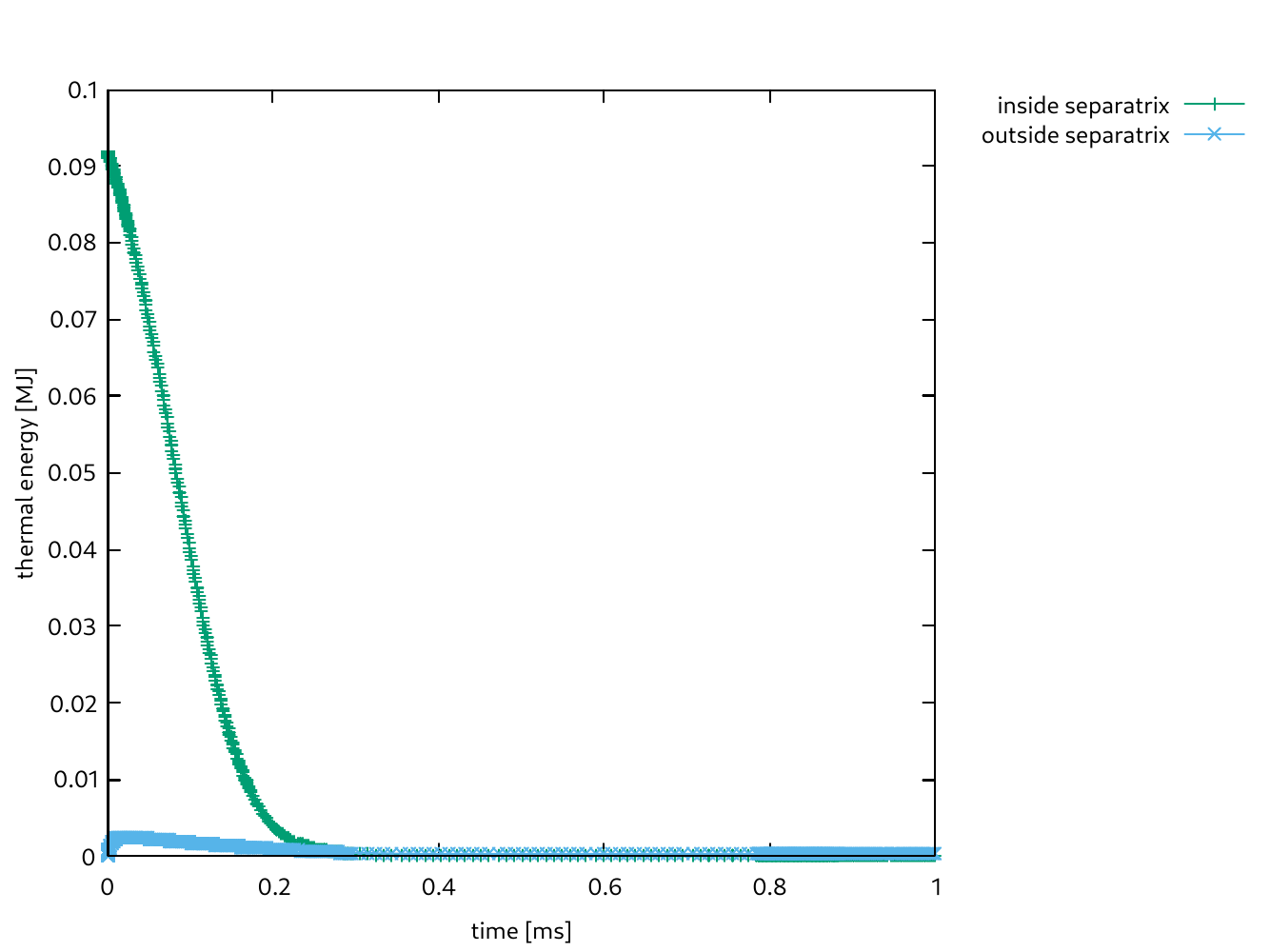}
        \caption{Thermal energy}
        \label{fig:vacther}
    \end{subfigure}
    \begin{subfigure}[b]{0.49\textwidth}
        \centering
        \includegraphics[width=1.1\textwidth]{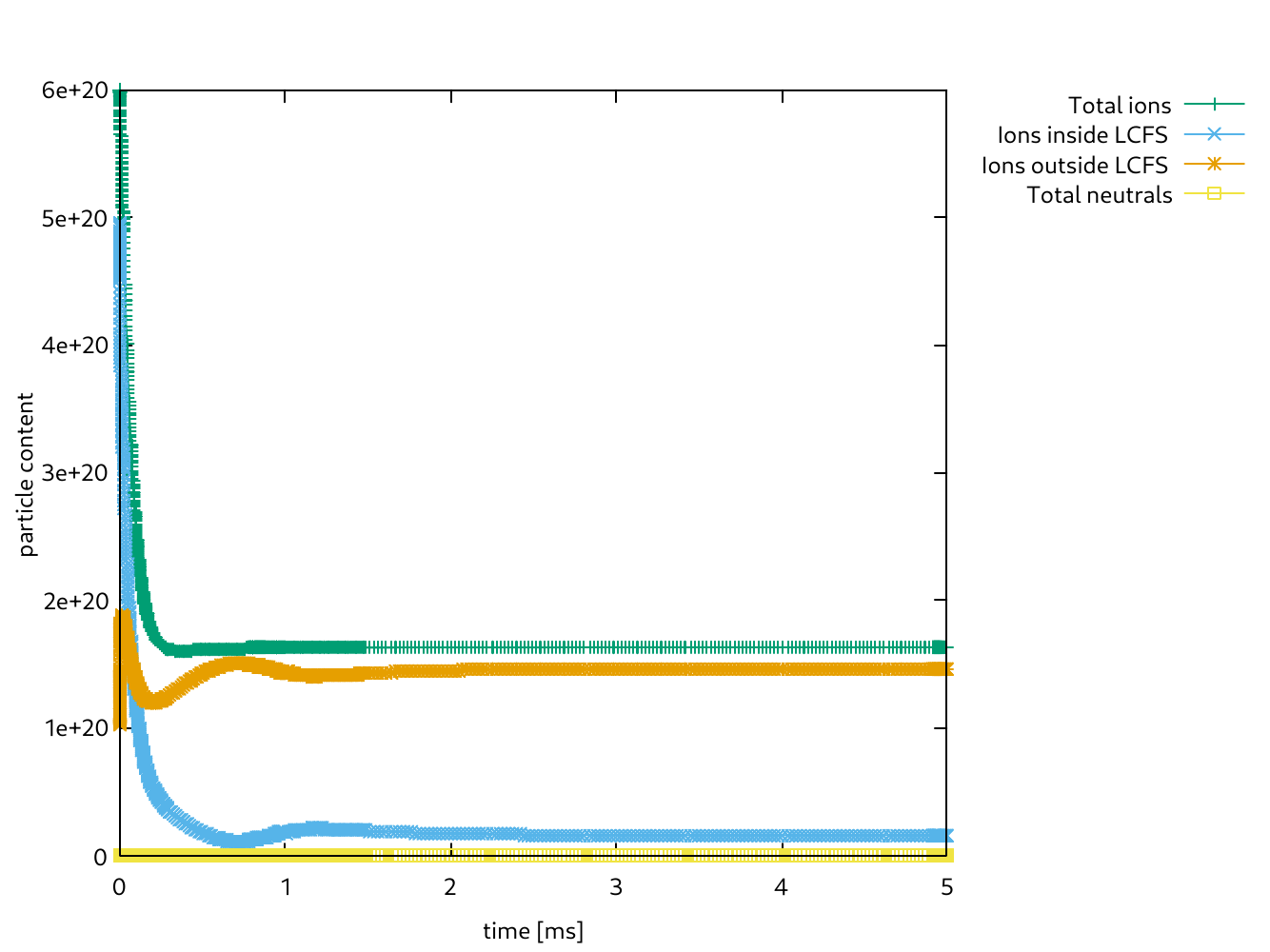}
        \caption{Particle content}
        \label{fig:vacpart}
    \end{subfigure}
    \vspace{\baselineskip} 
    \begin{subfigure}[b]{0.49\textwidth}
        \centering
        \includegraphics[width=1.1\textwidth]{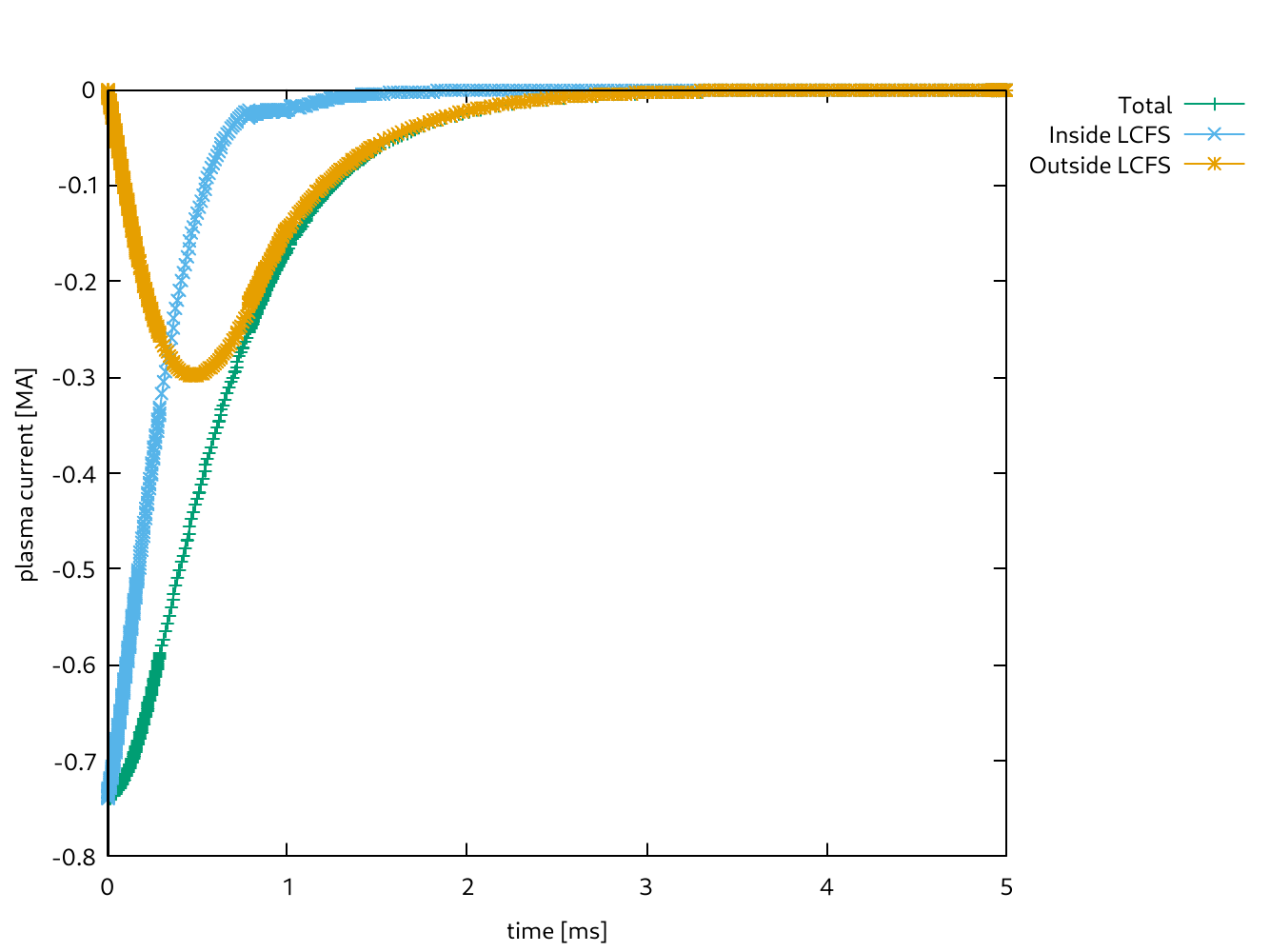}
        \caption{Plasma current}
        \label{fig:vaccurr}
    \end{subfigure}

    \caption{Time evolution of different plasma parameters from their initial conditions into a vacuum state. The abbreviation LCFS denotes the last closed flux surface}
    \label{fig:vacuum}
\end{figure}

The time evolution of the plasma current, particle number and thermal energy are shown in figure \ref{fig:vacuum} until an approximate vacuum state is reached. Due to numerical reasons, the simulation was initially run with only its axisymmetric component before including the n=1,…,6 harmonics.\\

\section{Plasma simulations}
This report will also briefly treat plasma simulations to investigate the screening effect within plasmas. The most relevant parameters are shown in table \ref{tab:plasmavars}. The parameter for the diamagnetic drift was selected to realistically model the background plasma flows, enabling the simulation of rotational screening and diamagnetic background flows. Due to time constraints, the plasma simulations were performed using only the toroidal modes n=0,1,2.

\chapter{Results}
\label{chap:results}

The analysis that was performed in this report consists of three parts. First, CARIDDI will be benchmarked against STARWALL using an RMP field that is static over time. Next, the investigation will focus on the screening of RMP fields by conducting volumetric structures in CARIDDI. This will involve the application of a current through the RMP coils that oscillates at various frequencies, considering both scenarios with and without the presence of a passive stabilizing loop (PSL) and a wall. Finally, a brief investigation of the induced eddy currents in a plasma resulting from the ramp-up phase of an n=2 RMP will be presented. The ramp-up phase is the time period during which the coil currents build up to their target amplitude.\\

\section{Steady coil current}
\begin{figure}[!ht]
     \centering
         \includegraphics[width=1.05\textwidth]{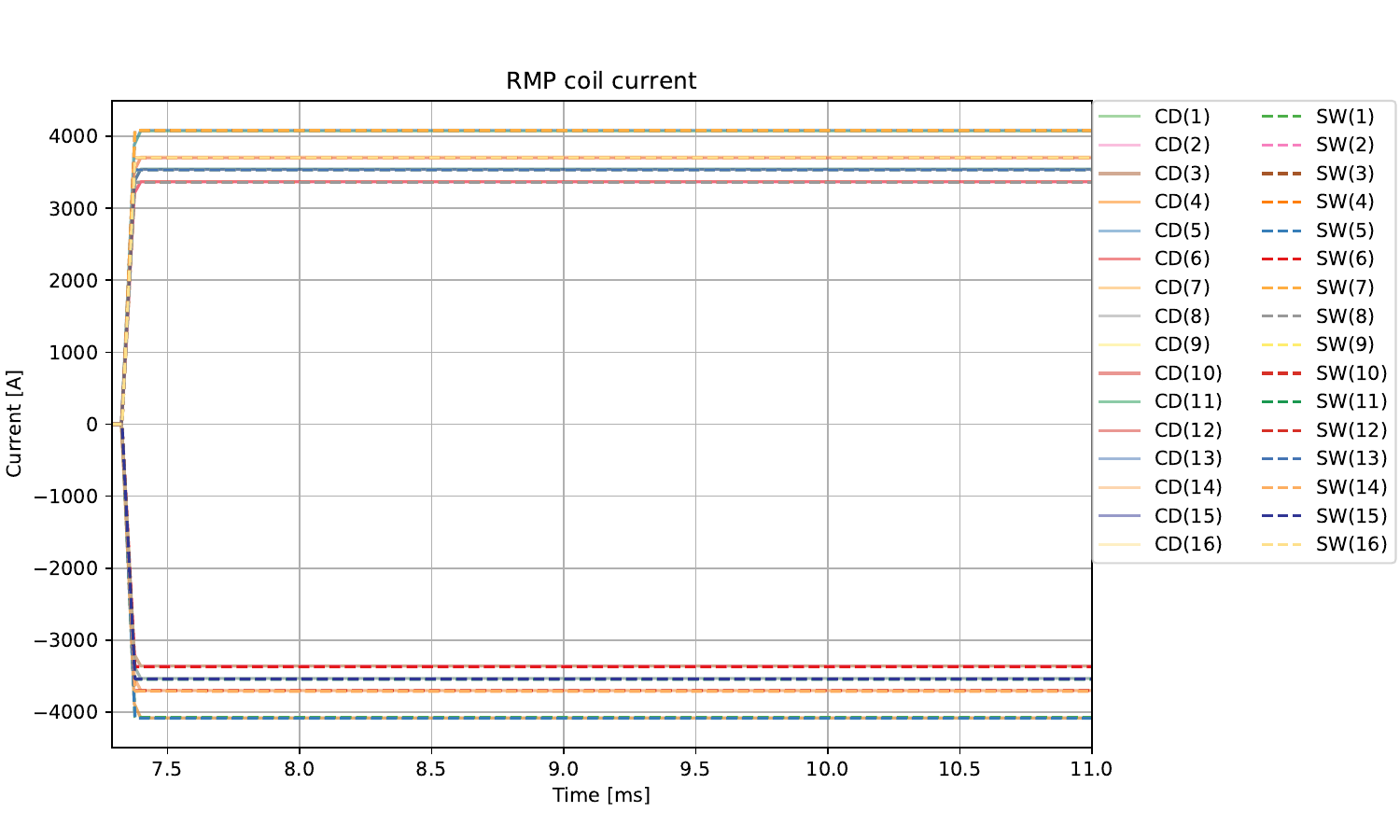}
        \caption{RMP coil currents in STARWALL and CARIDDI as a function of time for a static n=2 coil configuration. CD(i) and SW(i) correspond to the i-th RMP coil in CARIDDI and STARWALL, respectively}
        \label{fig:coilcurrents_stat}
\end{figure}

Figure \ref{fig:coilcurrents_stat} displays the time evolution of the RMP coil currents in the case of a static perturbation. The sign of the coil currents was reversed in CARIDDI, because the definition of the positive current direction is different as compared to STARWALL. 

\begin{figure}[!ht]
     \centering
     \begin{subfigure}[b]{0.47\textwidth}
         \centering
         \includegraphics[width=1.15\textwidth]{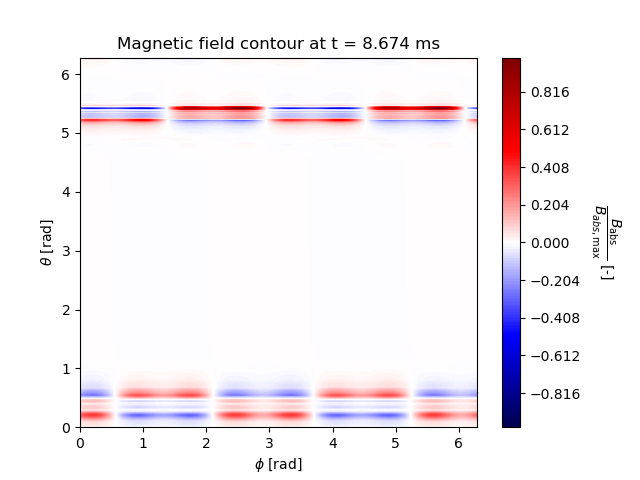}
         \caption{STARWALL}
         \label{fig:BabsStatSW}
     \end{subfigure}
     \hfill
     \begin{subfigure}[b]{0.47\textwidth}
         \centering
         \includegraphics[width=1.15\textwidth]{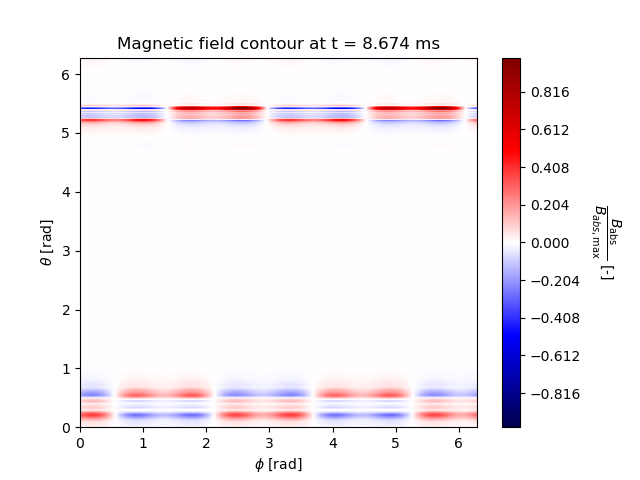}
         \caption{CARIDDI}
         \label{fig:BabsstatCD}
     \end{subfigure}
        \caption{Figures displaying the renormalized magnetic field at the computational boundary at a fixed time for simulations using STARWALL and CARIDDI, respectively. $\phi$ denotes the toroidal direction and $\theta$ the poloidal direction}
        \label{fig:B_abs_stat}
\end{figure}

The (static) magnetic fields that are the result of the applied RMP coil currents are shown in figure \ref{fig:B_abs_stat}. The field magnitudes in the figures were normalized using the maximum value of the magnetic field. 

The magnetic fields simulated by STARWALL and CARIDDI are a good match, establishing a good benchmark for CARIDDI. Small variations are likely caused by slightly different coil geometries. In \ref{sec:Axtra}, plots can be found that describe the time evolution of both the magnetic field at a fixed point and of the maximum value of the magnetic field, which show that the magnetic fields are constant over time. Furthermore, it was found that the maximum magnitude of the magnetic field in CARIDDI is roughly 20$\%$ higher than in STARWALL. The observed magnitude difference is the result of the different representations of the RMP coils between the STARWALL and CARIDDI response files, as discussed in section \ref{sec:SWCD}. \\
Section \ref{sec:Axtra} also contains a heatmap that shows the differences between plots \ref{fig:BabsStatSW} and \ref{fig:BabsstatCD} (i.e. $B_{abs, STARWALL}-B_{abs, CARIDDI}$).\\

\begin{figure}[!ht]
    \centering
    \includegraphics[width=0.7\textwidth]{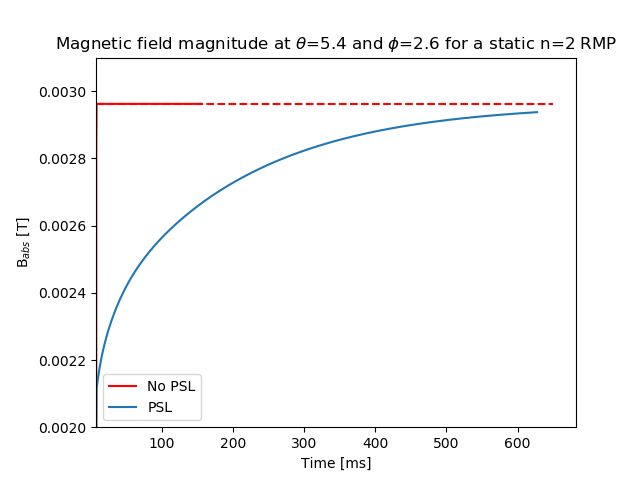}
    \caption{The time evolution of the magnetic field magnitude after the coil current ramp up in CARIDDI at a given point on the boundary, with and without wall and PSL. The dotted line was extrapolated from the simulation without PSL}
    \label{fig:Sat}
\end{figure}

Figure \ref{fig:Sat} shows the magnetic field that is produced by the static coil currents in CARIDDI, with and without PSL and wall. The amplitude of the magnetic field in the presence of a wall and PSL rises until it reaches the same value of the magnetic field without PSL. This effect is attributed to the eddy currents induced by the time-varying magnetic field during the ramp-up phase. It takes approximately 630 milliseconds for all eddy currents to fully dissipate from the wall and PSL, resulting in a magnetic field similar to the unscreened case. 

\section{Time-varying coil current}
As mentioned in chapter \ref{chap:background}, the screening by conducting structures occurs as a result of a time varying magnetic field. The RMP coil current was modified according to the formula:

\begin{equation}
    I_i(t) = I_{0,i}\cdot(0.5-0.5\cdot\tanh(-(t-c_1)/c_2))\cdot \sin(\hat{\omega}\cdot t)
\end{equation}

where $I_{0,i}$ represents the current of coil $i$ for the static $n=2$ RMP (see table \ref{tab:coilcurrs}), $t$ denotes the Alfvén time, $\hat{\omega}$ signifies the radial frequency (in Alfvén units), and $c_1$ and $c_2$ are constants that influence the ramp-up of the initial oscillation, given by $c_1=0.3\cdot f$ and $c_2=\frac{10}{f}$, where $f$ is the coil current oscillation frequency in Hz. The hyperbolic tangent ($\tanh$) function is used to achieve a gradual ramp-up, enhancing the stability of the simulation during the initial phase of the coil current.\\

\begin{figure}[!ht]
    \centering
    \includegraphics[width=0.8\textwidth]{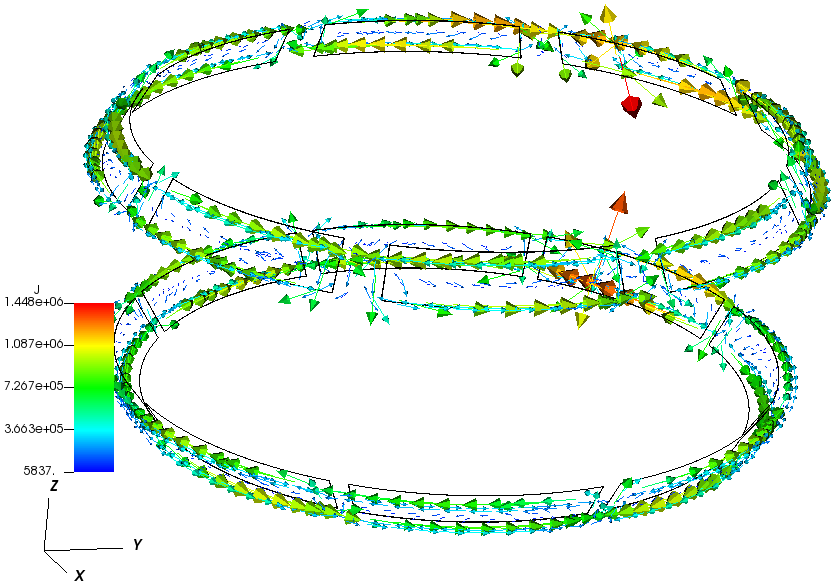}
    \caption{The eddy currents that are induced in the PSL as a result of a time varying RMP for a coil current oscillation frequency of 1 kHz. The black lines correspond to the boundary shapes of the RMPs. The colorbar represents the current density in $A\cdot m^{-2}$ }
    \label{fig:Eddy}
\end{figure}

\begin{figure*}[!ht]
    \centering

    \begin{subfigure}[b]{0.45\textwidth}
        \centering
        \includegraphics[width=1.35\textwidth]{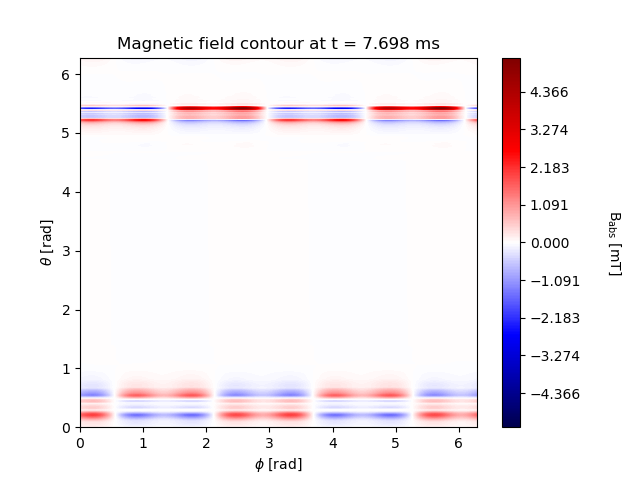}
        \caption{Static coil current}
        \label{fig:statPSL}
    \end{subfigure}%
    \hfill
    \begin{subfigure}[b]{0.45\textwidth}
        \centering
        \includegraphics[width=1.35\textwidth]{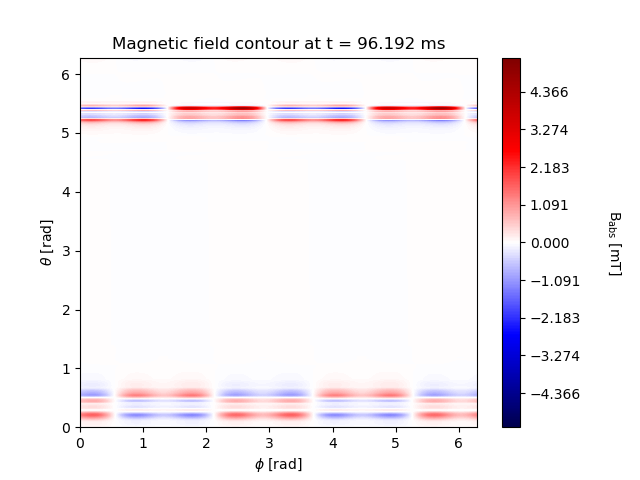}
        \caption{3 Hz}
        \label{fig:3HzPSL}
    \end{subfigure}

    \vspace{\floatsep}  

    \begin{subfigure}[b]{0.45\textwidth}
        \centering
        \includegraphics[width=1.35\textwidth]{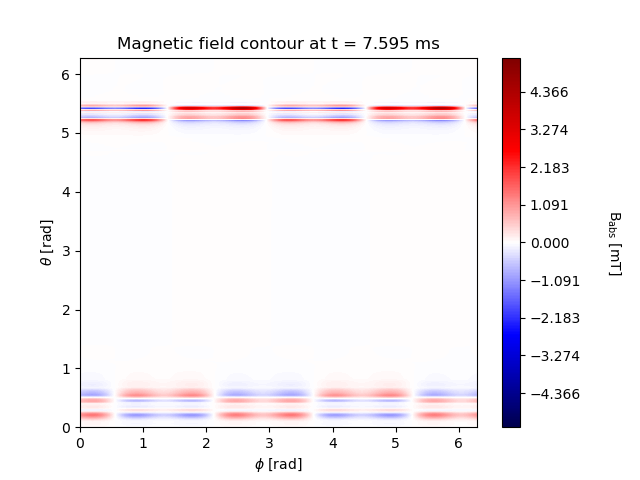}
        \caption{1 kHz}
        \label{fig:1kHzPSL}
    \end{subfigure}%
    
    \caption{Magnetic field magnitude at the computational boundary for coil currents oscillating at different frequencies in the presence of a wall and PSL. The chosen time corresponds to the moment of the peak of the coil current oscillation}
    \label{fig:fullfig}
\end{figure*}

The analysis was performed for oscillation frequencies ranging from 3 Hz up until 10 kHz, using a CARIDDI response file with and without PSL and wall. Figure \ref{fig:Eddy} displays the induced eddy currents in the PSL as the result of the applied RMP field. The resulting magnetic field heatmaps are shown in figure \ref{fig:fullfig} for different frequencies. The magnetic fields have the same shape, but different magnitude, which is the result of the screening by the PSL and wall. \\


\begin{figure*}[!ht]
    \centering

    \begin{subfigure}[b]{0.45\textwidth}
        \centering
        \includegraphics[width=1.33\textwidth]{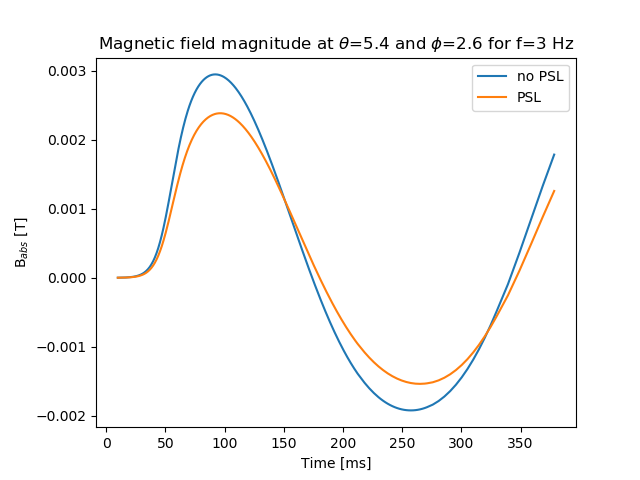}
        \caption{3 Hz}
        \label{fig:3HzBscreen}
    \end{subfigure}%
    \hfill
    \begin{subfigure}[b]{0.45\textwidth}
        \centering
        \includegraphics[width=1.33\textwidth]{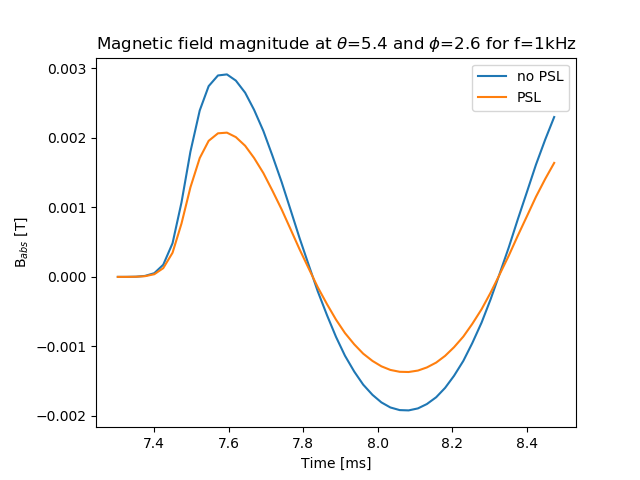}
        \caption{1 kHz}
        \label{fig:1kHzBscreen}
    \end{subfigure}

    \caption{Time evolution of the magnetic field magnitude at a fixed point on the boundary for different coil current oscillation frequencies. Note that this is the initial phase of the oscillation, i.e. including the slow ramp-up}
    \label{fig:bscreen}
\end{figure*}

\begin{figure}[!ht]
    \centering
    \includegraphics[width=0.7\textwidth]{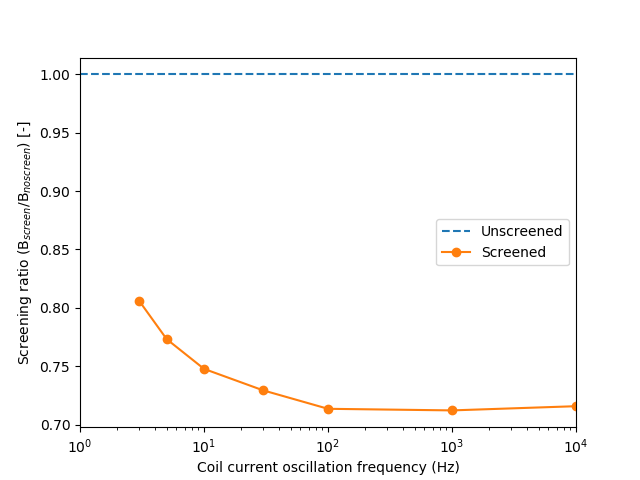}
    \caption{The screening ratio as a function of coil current oscillation frequency, calculated at the peak magnetic field of one full oscillation. All values were calculated for the same point at the boundary.}
    \label{fig:bscreens}
\end{figure}

Figures \ref{fig:3HzBscreen} and \ref{fig:1kHzBscreen} show the time evolution of the magnetic field magnitude at a fixed point on the boundary for coil current frequencies of 3 Hz and 1 kHz, with and without wall and PSL. The presence of a wall and PSL results in a notably lower magnetic field magnitude, indicating screening effects in CARIDDI. Figure \ref{fig:bscreens} shows the screening fraction for all simulated frequencies. The screening fraction is defined as $\frac{|\textbf{B}_{PSL}|}{|\textbf{B}_{no PSL}|}$. At coil current oscillation frequencies below 100 Hz, the screening fraction matches qualitatively well with the behaviour expected from figure \ref{fig:freqscreen}, showing increasing screening with rising frequency. The screening saturates at frequencies above 100 Hz. This behaviour results from a slower decay of the eddy currents in the PSL compared to the RMP oscillation time scale (inverse of the RMP frequency). It is suspected that the spatial resolution of the PSL in our simulation is insufficient to describe the screening amplitude accurately. A higher resolution is expected to lead to a similar dependency, but with a more complete screening at high frequencies. \\
\\ 




\newpage

\section{Plasma screening}
\label{sec:plasma}
\begin{figure*}[!ht]
    \centering

    \begin{subfigure}[b]{0.47\textwidth}
        \centering
        \includegraphics[width=1.3\textwidth]{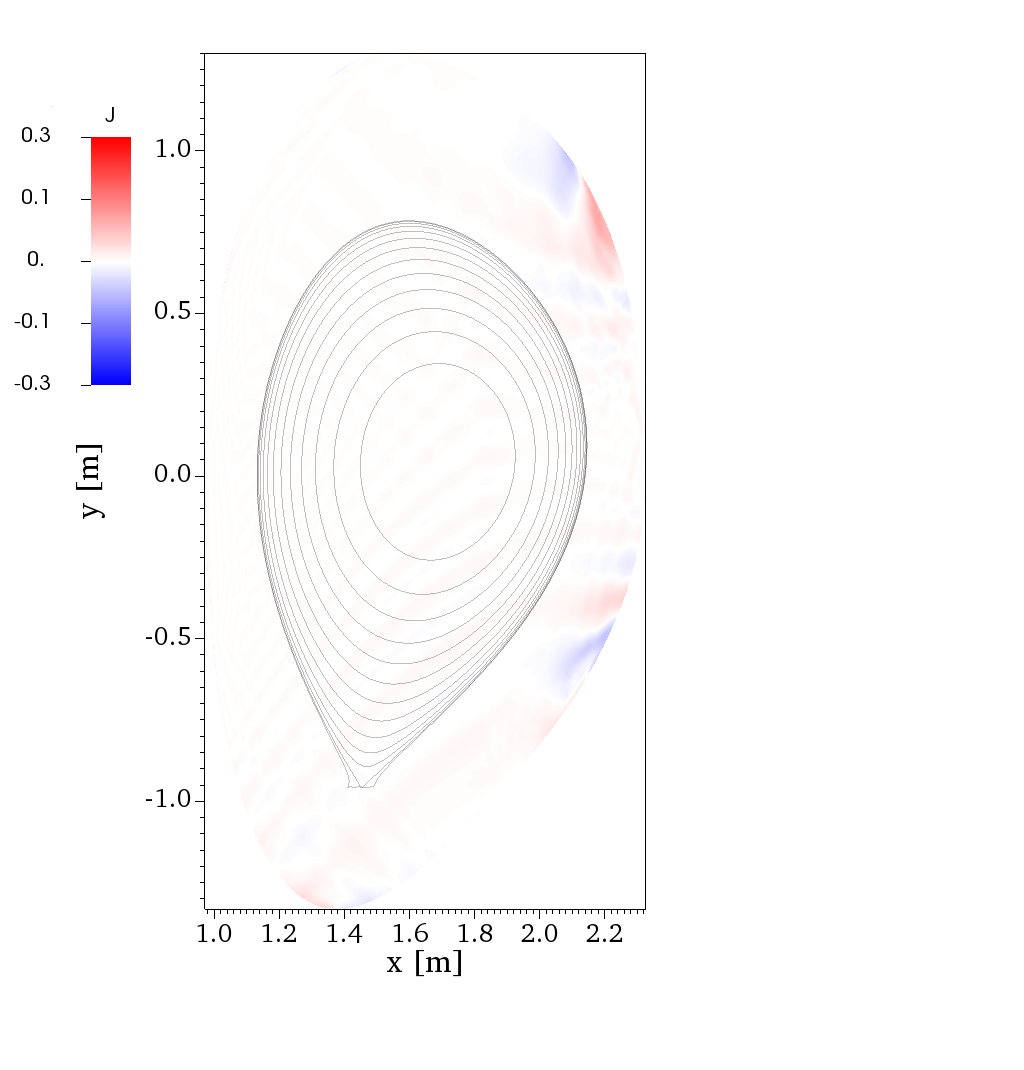}
        \caption{Vacuum}
        \label{fig:JvacuumCross}
    \end{subfigure}%
    \hfill
    \begin{subfigure}[b]{0.47\textwidth}
        \centering
        \includegraphics[width=1.3\textwidth]{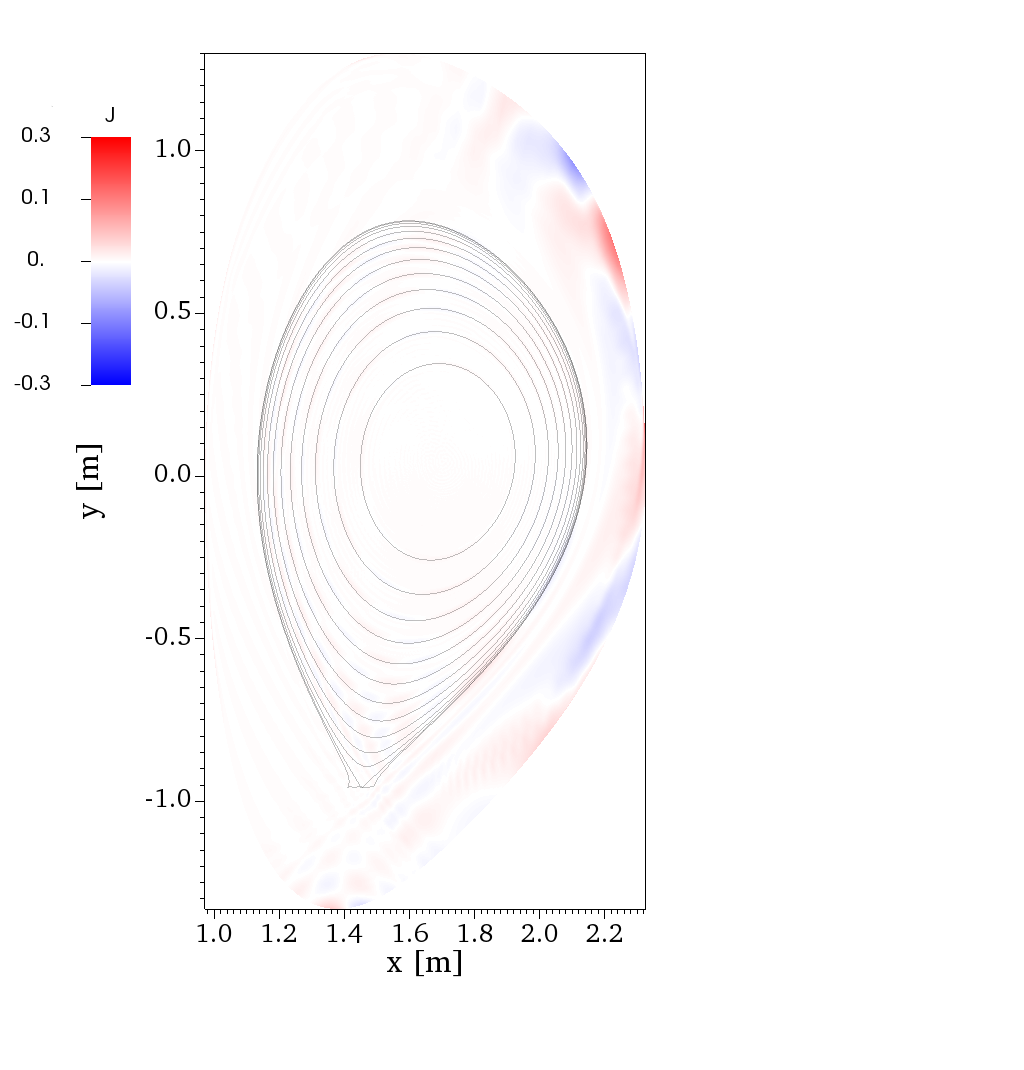}
        \caption{Plasma}
        \label{fig:JplasmaCross}
    \end{subfigure}

    \caption{Poloidal cross-section illustrating the current density directly after the ramp-up of the RMP coil current for an n=2 static RMP in the presence of either a vacuum or a plasma. The plots exclude the axisymmetric plasma component, the unit of the current density is $[A m^{-2}]$, and the black lines represent the n/2 flux surfaces}
    \label{fig:Jpscreen}
\end{figure*}

\begin{figure*}[!ht]
    \centering

    \begin{subfigure}[b]{0.47\textwidth}
        \centering
        \includegraphics[width=1.3\textwidth]{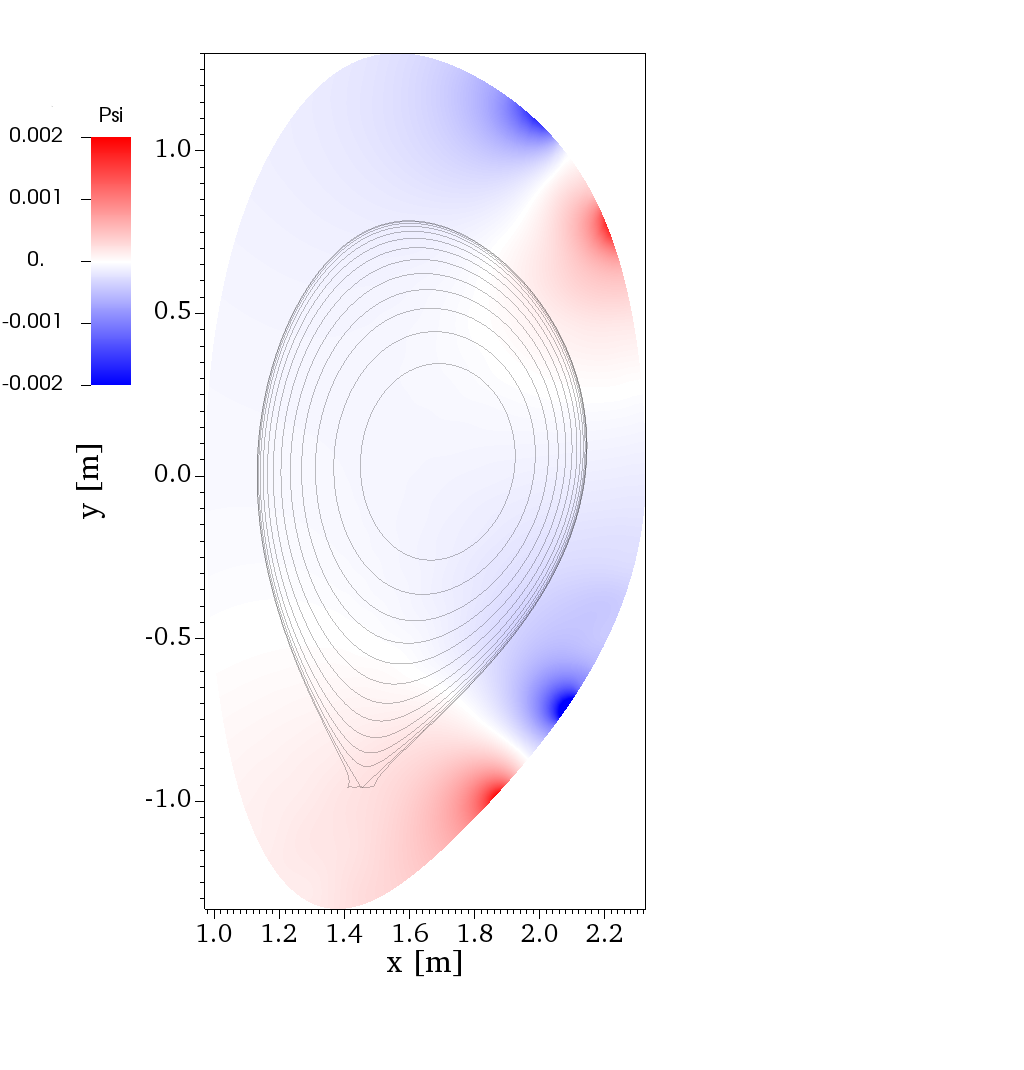}
        \caption{Vacuum}
        \label{fig:AvacuumCross}
    \end{subfigure}%
    \hfill
    \begin{subfigure}[b]{0.47\textwidth}
        \centering
        \includegraphics[width=1.3\textwidth]{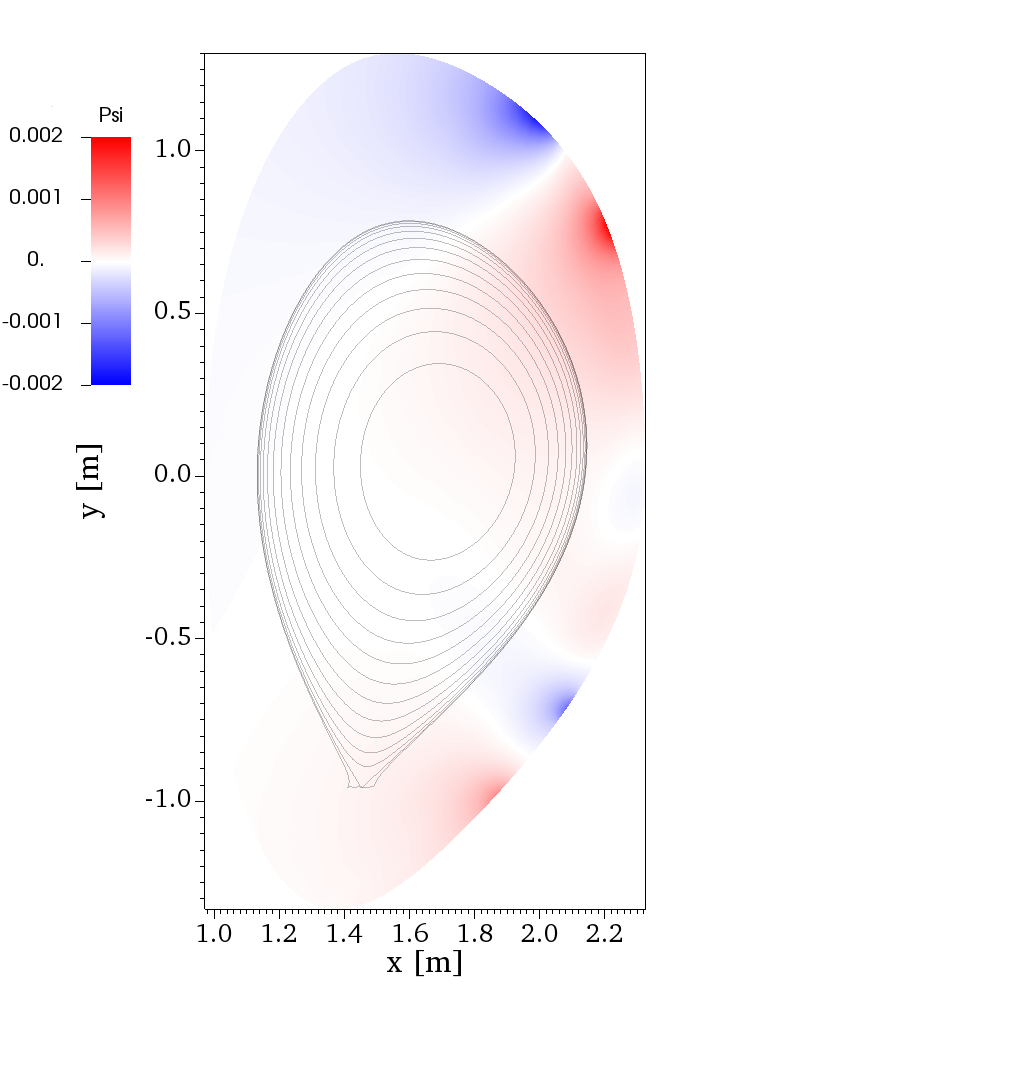}
        \caption{Plasma}
        \label{fig:AplasmaCross}
    \end{subfigure}

    \caption{Poloidal cross-section illustrating the poloidal magnetic flux directly after the ramp-up of the RMP coil current for an n=2 static RMP in the presence of either a vacuum or a plasma.  The plots exclude the axisymmetric plasma component, the unit of the poloidal magnetic flux is $[Tm^2]$, and the black lines represent the n/2 flux surfaces}
    \label{fig:Apscreen}
\end{figure*}

Since a plasma is conductive, the RMPs can also induce eddy currents within the plasma, locally screening the perturbation. Figure \ref{fig:Jpscreen} displays the current density in the presence of either a vacuum or a plasma immediately after the ramp-up of the RMP coil current. Noticeable structures appear in the plasma that are absent in the vacuum case, suggesting the induction of eddy currents. A finite amplitude of currents is also visible in the “vacuum” case, which is a result of solving the same MHD equations but at lower temperature and thus higher resistivity. The resistivity in this quasi-vacuum is not high enough to prevent current induction in the domain entirely. Note that the plasma screening results from the combined effects of eddy currents induced in the plasma by the ramp-up of the RMP field and the rotational screening of electrons with an ExB drift. \\ The effect of these screening currents is reflected in the resulting perturbation of the poloidal magnetic flux, as shown in the slight differences between figures \ref{fig:AvacuumCross} and \ref{fig:AplasmaCross}.  In the plasma case, the magnitude and penetration depth of the applied RMP are reduced compared to the vacuum case. This qualitative result shows an effect that is an important consideration when implementing RMPs for ELM suppression.  \\

\begin{figure*}[!ht]
    \centering

    \begin{subfigure}[b]{0.5\textwidth}
        \centering
        \includegraphics[width=1.3\textwidth]{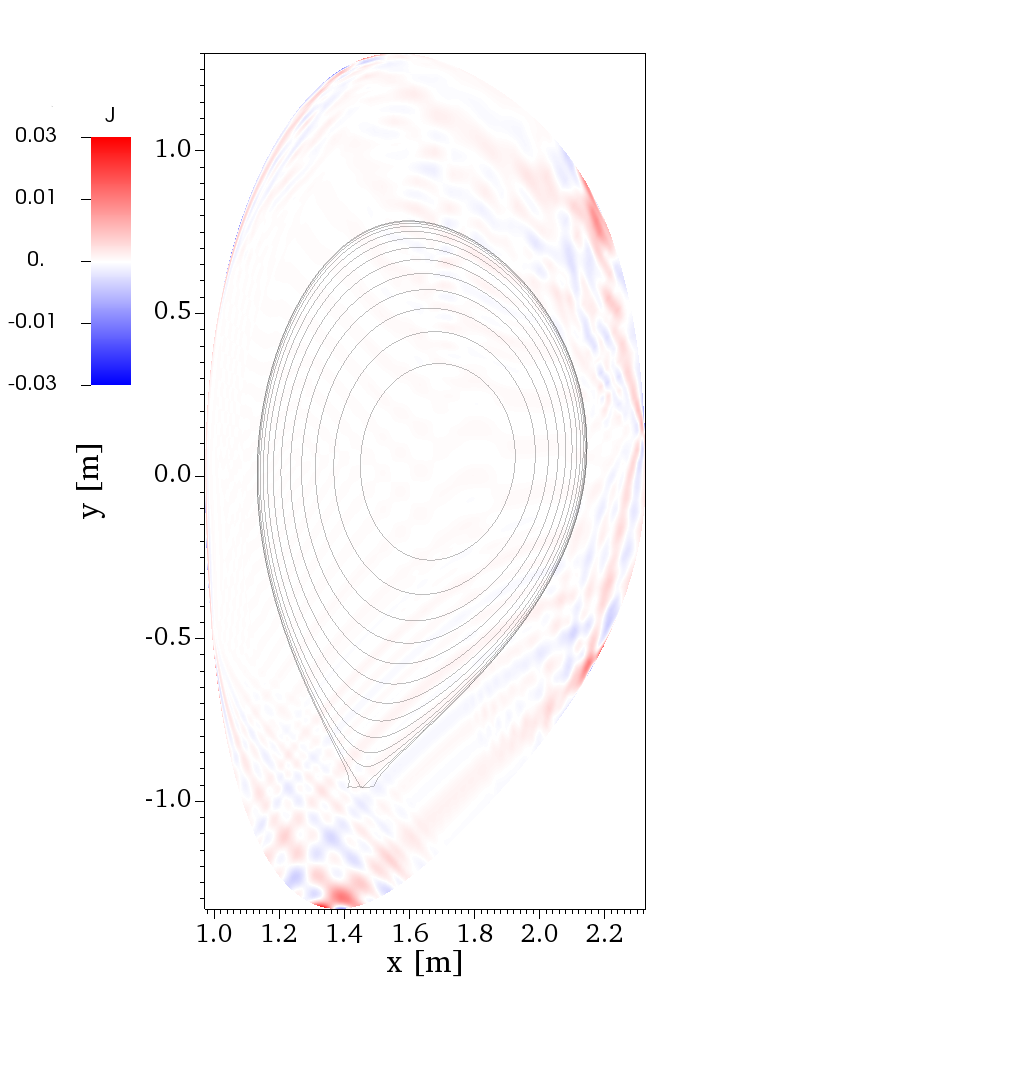}
        \caption{Vacuum}
        \label{fig:JvacuumCross_long}
    \end{subfigure}%
    \hfill
    \begin{subfigure}[b]{0.5\textwidth}
        \centering
        \includegraphics[width=1.3\textwidth]{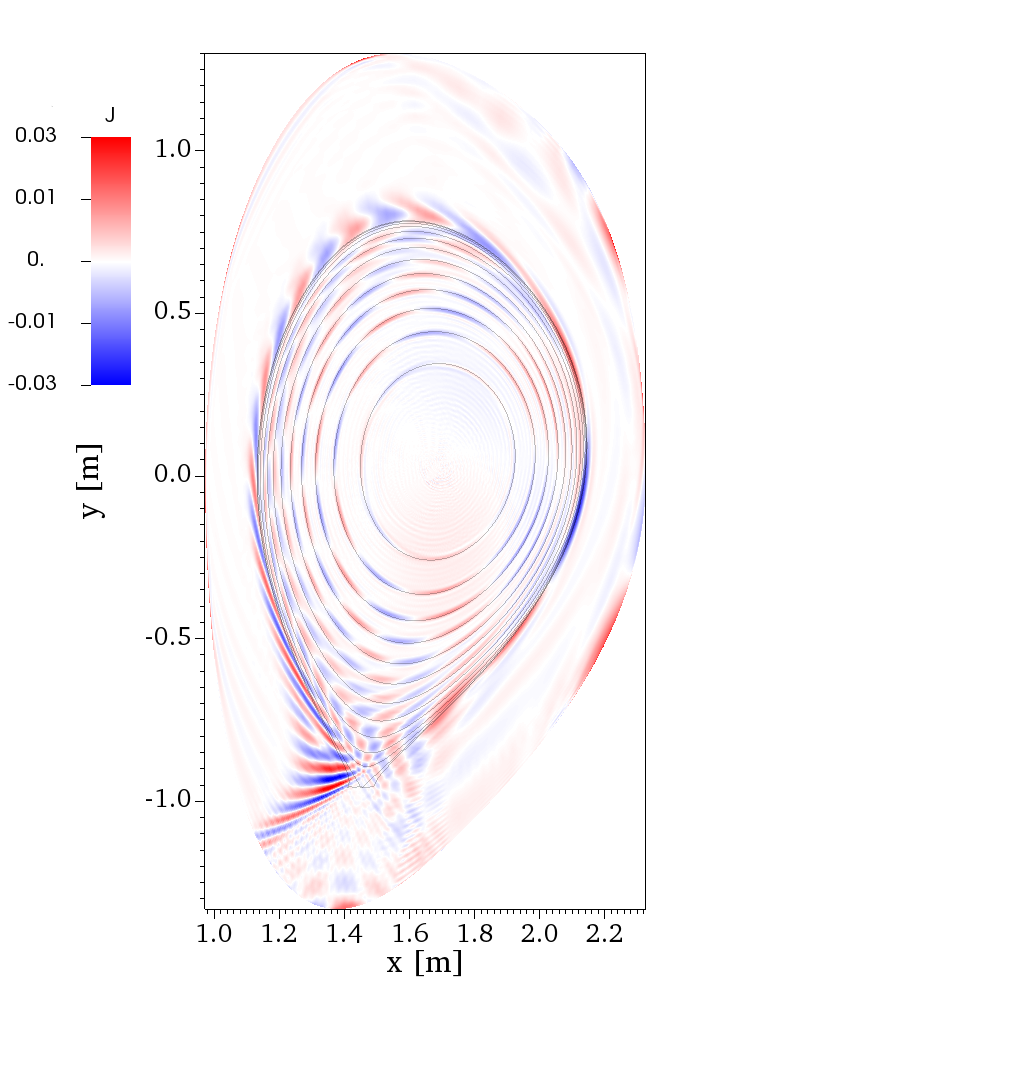}
        \caption{Plasma}
        \label{fig:JplasmaCross_long_tau_not0}
    \end{subfigure}%

    \caption{Poloidal cross-section illustrating the current density, a significantly long time after the ramp-up of the RMP coil current for an n=2 static RMP in the presence of either a vacuum or a plasma.  The plots exclude the axisymmetric plasma component, the unit of the current density is $[A m^{-2}]$, and the black lines represent the n/2 flux surfaces}
    \label{fig:Long_Jpscreen}
\end{figure*}

Figure \ref{fig:Long_Jpscreen} displays the current density at a significantly extended time after the ramp-up of the RMP coils. Here, the differences between the plasma and vacuum case are very evident. The plasma case shows plasma current density structures at the resonant surfaces inside the plasma, whereas the vacuum case displays no such structures.\\



\chapter{Discussion }
\label{chap:discussion}
 As discussed in chapter \ref{chap:results},  the simulations presented in this report show qualitative agreement with the behaviour expected from figure \ref{fig:freqscreen}, but didn't show agreement for coil current oscillation frequencies above 100 Hz. It is crucial to note that Figure \ref{fig:freqscreen} considers the coil casing of the RMP coils, resulting in a more pronounced decline at frequencies above 1 kHz. The results presented in this report don't include this casing, and thus the screening fraction in figure \ref{fig:bscreens} saturates at high frequencies.  Therefore, the coil casing should be taken into account or compensated for when aiming for exact quantitative agreement. \\
 Furthermore, it is recommended to repeat the investigation with simulations featuring higher spatial resolution for the conducting structures. This investigation will likely show more complete screening at coil current oscillation frequencies between 100 and 1000 Hz.
 
 The analysis of the plasma screening showed the qualitative occurrence of plasma screening, but did not go in much detail due to time constraints. It is recommended to explore the specific impact of the plasma by comparing various plasma densities for a more comprehensive understanding. Additional recommendations for further research include; 

 \begin{itemize}
    \item Investigating plasma screening for time-varying RMP coil currents
    \item  Analysing the individual effect of diamagnetic electron rotation on the total plasma screening
     \item Repeating the analysis using a grid with an x-point configuration to better resemble ASDEX Upgrade (AUG)
     \item  Exploring different RMP modes (i.e., n=1, 4, 8) in the analysis to assess their impact on the intensity of the screening effect
     \item  Considering alternative configurations of conducting structures, such as those from DIII-D, ITER, or other fusion tokamaks
     \item Repeating the analysis with a bigger grid size
 \end{itemize}

 All outcomes presented in this report were generated using JOREK, a thoroughly validated MHD simulation code. The analysis in STARWALL and CARIDDI both used identical initial simulations for the vacuum state, ensuring consistency and eliminating potential variations. The response files, responsible for modelling the behaviour of surrounding conducting structures, were prepared by collaborators that have a comprehensive understanding of the respective codes, minimizing the likelihood of mistakes in the response files. Furthermore, the desired properties of the modelled structures were clearly communicated, and the same grid was used for all simulations. While some bugs were encountered during the simulations, they were addressed by adapting the simulation methods.
 

{\let\clearpage\relax \chapter{Conclusion and outlook} \label{chap:conclusion}}

The main objective for the investigation presented in this report was to characterize the screening of magnetic perturbation fields produced by RMP coils by passive conducting structures using the volumetric resistive wall code CARIDDI. 

At first, a benchmark between the STARWALL and CARIDDI codes was carried out for the RMP induced magnetic fields, which showed good agreement. Remaining differences are expected to result from the different geometries of the coil models in the two codes (thin-wall versus volumetric). After this verification step, simulations with time-varying RMP fields oscillating at different frequencies were carried out, which show partial screening by the PSL structures. As expected, the screening has lower intensity at low frequencies. An additional screening at frequencies above 1 kHz that was observed in earlier engineering studies could not be recreated, as the coil housing responsible for this screening was not incorporated in the CARIDDI coil model used here. Thus, while the general expected trends of the screening are qualitatively recovered, in particular the absolute amplitude of the screening (saturating at ~28\% in the simulations carried out here) is lower than expected. This is likely a result of an insufficient resolution of the PSL model. An additional test with a different PSL resolution was not possible within the timescale of this internship.\\
 
Besides these studies, which concentrate on the RMP screening by conducting structures and thus were using a vacuum-like setup inside the JOREK computational domain, also a few tests were done with a plasma configuration. As expected, the conducting and rotating plasma can partially screen the RMP fields as well, by induced eddy currents. The induction of such eddy currents at rational surfaces inside the plasma was demonstrated.\\
 
Altogether, this study demonstrated and verified the capability of JOREK-CARIDDI to model magnetic perturbation fields induced by 3D coils and showed that the screening of such fields by conducting structures and the plasma itself can be qualitatively captured. Quantitative verification for the screening by the PSL remains future work after a careful resolution scan.\\
 
This work contributes to fusion research in several relevant ways. First, RMP fields can play an important role for the suppression of large ELMs. Secondly, perturbation fields are also considered a possible method for the suppression of relativistic runaway electrons during disruptions. Screening plays an important role here for the time scales on which a sufficient amplitude of perturbation fields for such control purposes can be achieved. Furthermore, the ability to capture the effect of conducting structures onto magnetic fields also plays an important role for diagnostic purposes. The magnetic probes installed in fusion devices measure field perturbation induced by plasma instabilities, but screening effects by conducting structures modify the measurements. One to one comparisons between experimental measurements and virtual diagnostics applied to simulations are only possible if such effects are taken into account also in the computational studies.




\chapter*{Acknowledgements}
I would like to thank Matthias for welcoming me into his group and providing me with this opportunity to perform research with such an amazing team. Thank you for all the helpful discussions along the way and giving me so much time despite your overbooked agenda. An especially big thanks goes out to Verena for being my first line of defence against all problems that I encountered along the way, be it nasty errors, some theory that I didn't fully understand or dealing with the hideous interface of tightVNC. I would also like to thank Nina for her help in all things CARIDDI related; even though you were finishing your own PhD thesis, you still took the time to provide me with all the things I needed. Furthermore, I would also like to thank Nicola Isernia for his assistance with using CARIDDI. And of course, I also want to extend my warmest thanks to everybody in the JOREK group at the IPP. You all made me feel at home from my first day, aside from the occasional jokes about sandwiches at lunchtime. \\

I would also like to thank Roger for his assistance from the TU/e for being (in his own words) the 'well-informed outsider', and freeing up some time in his busy schedule. \\
Lastly, I would also like to thank Fusian dance crew for welcoming me into their group and by providing a welcome distraction from plasma physics.  \\

This research was carried out using funding from FuseNet and Erasmus+.

\bibliographystyle{biolett} 
\bibliography{bib}

\appendix
\chapter{Appendix}
\label{chap:appendix}


\section{JOREK input parameters}
\label{sec:params}

Table \ref{tab:JOREKparms} shows the parameters that were used to configure the JOREK build that was used in the analysis presented in this report. 

\begin{table}[ht]
\centering
\begin{tabular}{|c|c|c|}
\hline
Parameter & Value & Description \\
\hline
model & 600 & physics model\\
\hline
n\_tor	& 13 & number of toroidal harmonics \\
\hline
n\_period & 1 & toroidal periodicity\\
\hline
n\_plane & 24 & number of toroidal planes for real-space representation\\
\hline
\end{tabular}
\caption{The parameters that were used to set up the JOREK build that was used in this report}
\label{tab:JOREKparms}
\end{table}

Table \ref{tab:parmsvac} displays the most important parameters that were changed in JOREK to achieve the vacuum state.\\

\begin{table}[ht]
\centering
\begin{tabular}{|c|c|c|c|}
\hline
Parameter & Default JOREK value & New value & Description \\
\hline
eta & 1.e-5 & 1.e-4 & Resistivity at plasma center\\
\hline
eta\_T\_dependent	& .true. & .false. & Resistivity temperature dependent? \\
\hline
d\_perp & 1.e-5 & 1.e-3 & Coefficient for perpendicular particle diffusion profile \\
\hline
vk\_perp & 1.e-5 & 1.e-4 & Coefficient for perpendicular heat diffusion profile \\
\hline
visco & 1.e-5 & 8.3e-6 & Normalized viscosity at plasma center  \\
\hline
visco\_par & 1.e-5 & 8.3e-6 & Normalized parallel viscosity \\
\hline
visco\_T\_dependent &	.true.	& .false. & Viscosity temperature dependent?\\
\hline
keep\_current\_prof & .true . & .false. & Artificial source to keep the initial plasma current \\
\hline
zk\_par & 1 & 40000 & Parallel heat diffusion value in the plasma center  \\
\hline 
zk\_par\_max & 1.e20 & 4000& Maximum parallel heat diffusion value (due to numerical reasons) \\
\hline
\end{tabular}
\caption{The default and modified parameters in JOREK for the vacuum simulation case }
\label{tab:parmsvac}
\end{table}

Table \ref{tab:SWparams} displays the most important parameters that were used to generate the STARWALL response file, whereas table \ref{tab:CDparams} displays the most important CARIDDI input parameters to generate the CARIDDI response file.

\begin{table}[ht]
\centering
\begin{tabular}{|c|c|c|}
\hline
Parameter & Value & Description \\
\hline
i\_response & 2 & Calculates response matrix for resistive wall\\
\hline
n\_tor & 12 & number of toroidal modes\\
\hline
n\_period & 1 & toroidal periodicity \\
\hline
n\_points & 10 & number of grid points per boundary element\\
\hline
iwall & Fourier series (1) & representation of the wall \\
\hline
eta\_thin\_w & 1.e+02 & thin wall resistivity \\
\hline
rmp\_coil\_file & 'asdex\_rmp\_coils.nml' & location of the file that describes the RMP coil geometry \\
\hline
\end{tabular}
\caption{The STARWALL parameters used for calculating the response matrix }
\label{tab:SWparams}
\end{table}

\begin{table}[ht]
\centering
\begin{tabular}{|c|c|c|}
\hline
Parameter & Value & Description \\
\hline
ntori	& 10 & number of toroidal modes \\
\hline
nharmi & 1,2,3,4,5,6,7,8,9,10 & toroidal mode numbers\\
\hline
wall\_thickness\_J & 1.0e-4 & wall thickness\\
\hline
n\_points & 10 & number of grid points per boundary element\\
\hline
projection\_method\_j & trapz & representation of the wall \\
\hline
ETA(PSL) & 1.72e-08 & PSL resistivity (copper)\\
\hline
ETA(RMP coil) & 1.00e+00 & RMP coil resistivity \\
\hline
ETA(VV) & 0.73e-6 & vacuum vessel resistivity \\
\hline
\end{tabular}
\caption{The most important CARIDDI parameters used for calculating the response matrix }
\label{tab:CDparams}
\end{table}

Table \ref{tab:coilcurrs} contains the coil currents for each RMP coil in STARWALL for producing an n=2 perturbation. Note that the sign of the current was inverted for CARIDDI due to a different sign convention. 

\begin{table}[ht]
\centering
\begin{tabular}{|c|c|c|c|}
\hline
Coil (bottom row) & Current (A) & Coil (top row) & Current (A) \\
\hline
1	& 4073 & 9 & -3532 \\
\hline
2 & 3365 & 10 & 3700 \\
\hline
3 & -4081 & 11 & 3545\\
\hline
4 & -3359 & 12 & -3700\\
\hline
5 & 4084 & 13 & -3532 \\
\hline
6 & 3370 & 14 & 3711\\
\hline
7 & -4080 & 15 & 3540 \\
\hline
8 & -3360 & 16 & -3701 \\
\hline
\end{tabular}
\caption{The RMP coil currents that were used to produce a static n=2 perturbation }
\label{tab:coilcurrs}
\end{table}

\begin{table}[!ht]
\centering
\begin{tabular}{|c|c|c|c|}
\hline
Parameter & Default value & Simulation value & Description \\
\hline
eta & 1.e-5 & 3.e-8 & normalized resistivity at the plasma center \\
\hline
visco & 1.e-5 & 2.3e-8 & normalized viscosity at the plasma center\\
\hline
tauIC & 0 & 4.3522e-3 & scaling factor for diamagnetic terms \\
\hline
heatsource & 1.e-7 & 1.00e-6 & heat source amplitude\\
\hline
particlesource & 1e-5 & 1.50e-6 & particle source amplitude\\
\hline
edgeparticlesource & 0 & 1.00e-4 & edge particle source amplitude \\
\hline
\end{tabular}
\caption{The most relevant parameters that were modified for the plasma simulation presented in \ref{sec:plasma} }
\label{tab:plasmavars}
\end{table}

\newpage

\section{Extra plots}
\label{sec:Axtra}
This section contains supplementary plots to the results treated in chapter \ref{chap:results}. 

\subsection{Static coil current}

\begin{figure}[!ht]
    \centering
    \includegraphics[width=0.7\textwidth]{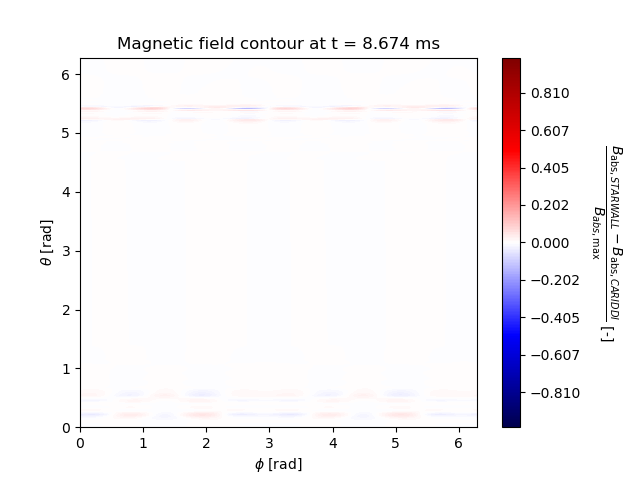}
    \caption{A plot illustrating the difference between the magnetic fields in STARWALL and CARIDDI}
    \label{fig:BabsStatComp}
\end{figure}

\begin{figure}[!h]
    \centering
    \includegraphics[width=0.7\linewidth]{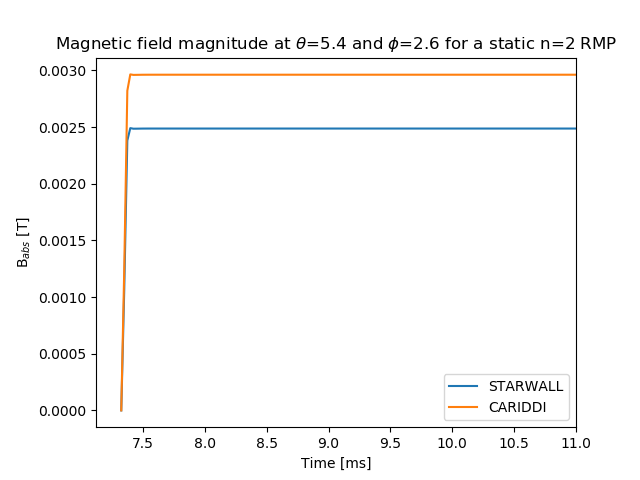}
    \caption{Magnetic field amplitude at a fixed point as a function of time for STARWALL and CARIDDI}
    \label{fig:statB}
\end{figure}

\newpage

\subsection{Oscillating coil current}

\begin{figure*}[!h]
    \centering

    \begin{subfigure}[b]{0.45\textwidth}
        \centering
        \includegraphics[width=1.33\textwidth]{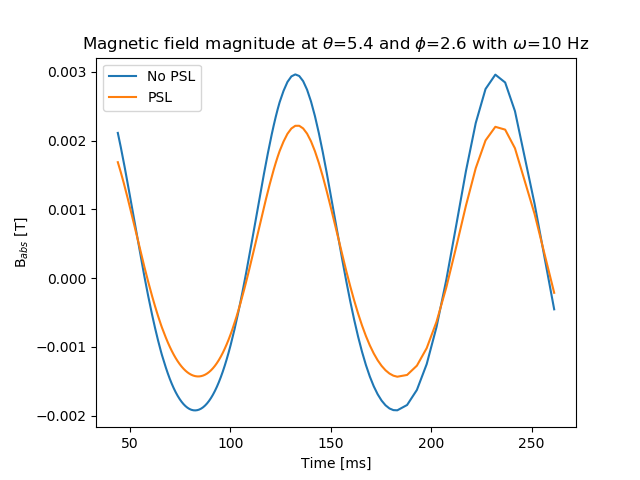}
        \caption{10 Hz}
        \label{fig:10HzBvalue}
    \end{subfigure}%
    \hfill
    \begin{subfigure}[b]{0.45\textwidth}
        \centering
        \includegraphics[width=1.33\textwidth]{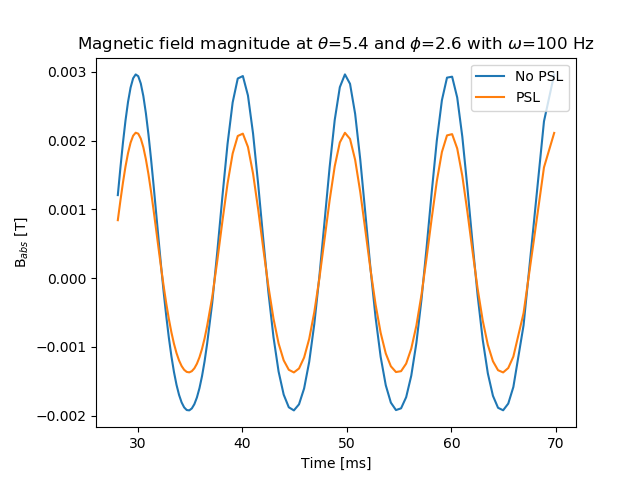}
        \caption{100 Hz}
        \label{fig:100HzBscreen}
    \end{subfigure}

    \caption{time evolution of the magnetic field amplitude at a fixed point on the boundary for different coil current oscillation frequencies}
    \label{fig:bscreenlow}
\end{figure*}

\begin{figure}[!h]
    \centering
    \includegraphics[width=0.7\linewidth]{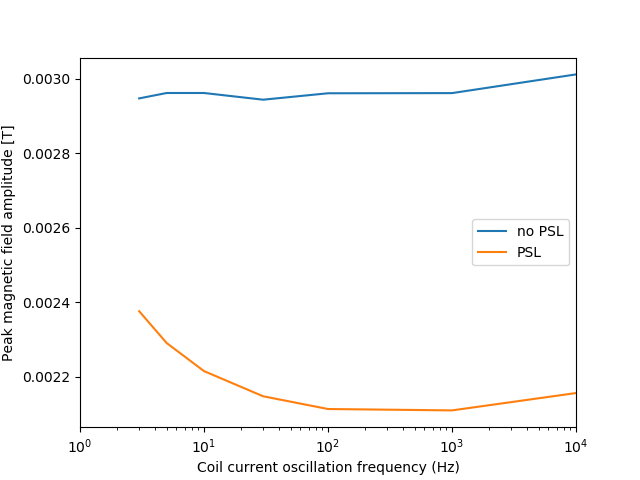}
    \caption{Magnetic field amplitude at a fixed point as a function of coil current oscillation frequency, both with and without the presence of conducting structures}
    \label{fig:allB}
\end{figure}

\begin{figure}
    \centering
    
    \begin{subfigure}{0.45\textwidth}
        \includegraphics[width=1.35\linewidth]{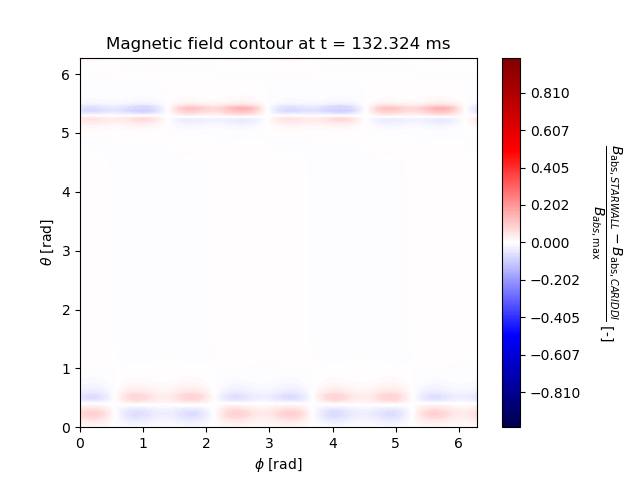}
        \caption{10Hz}
        \label{fig:cdiff10Hz}
    \end{subfigure}
    \hfill
    \begin{subfigure}{0.45\textwidth}
        \includegraphics[width=1.35\linewidth]{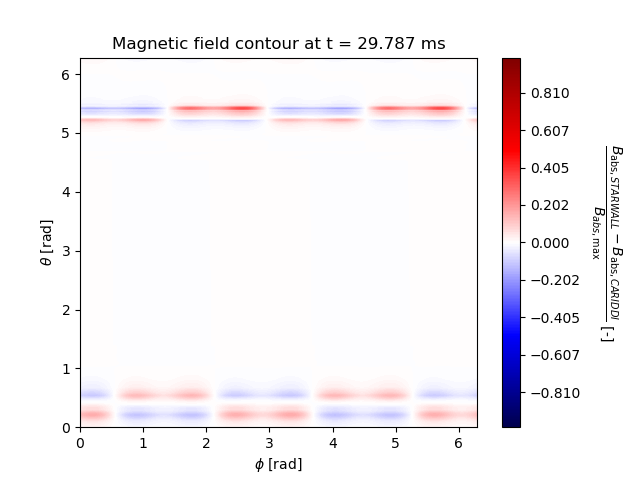}
        \caption{100 Hz}
        \label{fig:cdiff100Hz}
    \end{subfigure}

    \medskip

    \begin{subfigure}{0.45\textwidth}
        \includegraphics[width=1.35\linewidth]{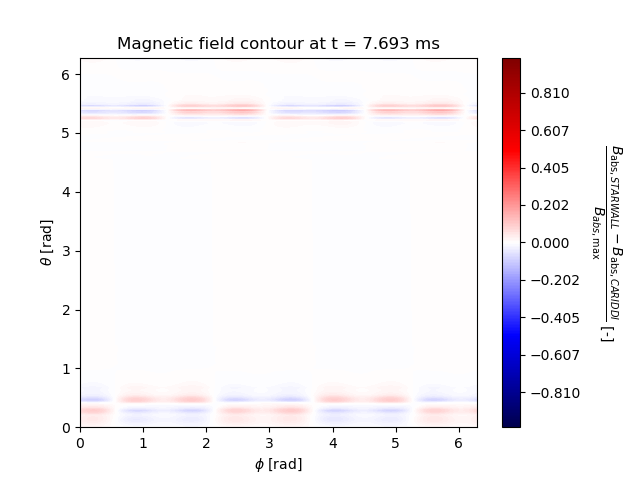}
        \caption{1 kHz}
        \label{fig:cdiff1kHz}
    \end{subfigure}
    \hfill
    \begin{subfigure}{0.45\textwidth}
        \includegraphics[width=1.35\linewidth]{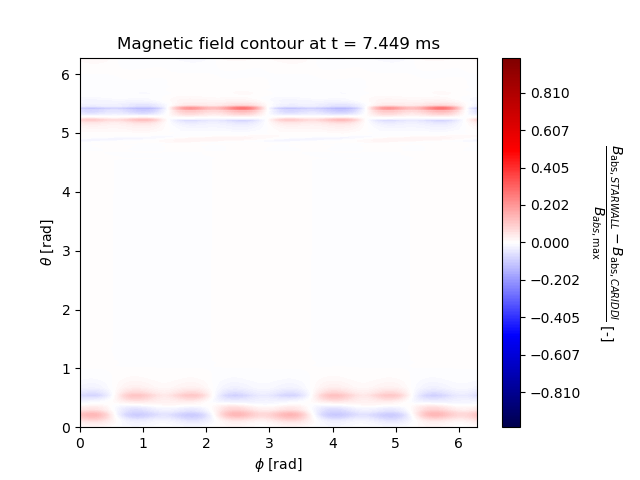}
        \caption{10kHz}
        \label{fig:cdiff10kHz}
    \end{subfigure}

    \caption{Heatmaps of the difference between the normalized magnitudes of the magnetic field at the computational boundary for different RMP oscillation frequencies}
    \label{fig:diffcontourfreqs}
\end{figure}

\end{document}